\documentclass{LMCS}
\usepackage{enumerate}
\usepackage{hyperref}
\usepackage{amsmath,amssymb,amsthm}
\usepackage{graphicx}
\usepackage[dvips]{epsfig}
\usepackage[T1]{fontenc}
\usepackage{fancyhdr,stmaryrd,color}
\usepackage{euscript}
\usepackage{url,xspace,array}
\usepackage[scaled=0.85]{luximono}

\DeclareMathAlphabet{\mathscr}{T1}{pzc}{m}{it}

\newtheorem{notation}[thm]{Notation}

\newcommand{\ie}{i.e., }
\newcommand{\fl}{\to}
\newcommand{\dfl}{\Rightarrow}
\newcommand{\tfl}{\Rrightarrow}
\newcommand{\ens}[1]{\left\{#1\right\}}
\newcommand{\mon}[1]{\left\langle#1\right\rangle}
\newcommand{\ul}[1]{\underline{#1}}
\newcommand{\abs}[1]{\left|#1\right|}
\newcommand{\norm}[1]{\abs{\abs{#1}}}
\newcommand{\tnorm}[1]{\abs{\abs{\abs{#1}}}}
\newcommand{\sem}[1]{\left\llbracket#1\right\rrbracket}
\newcommand{\roundup}[1]{\left\lceil#1\right\rceil}
\newcommand{\rounddown}[1]{\left\lfloor#1\right\rfloor}

\renewcommand{\phi}{\varphi}
\newcommand{\Cb}{\mathbb{C}}
\newcommand{\Nb}{\mathbb{N}}
\newcommand{\Zb}{\mathbb{Z}}
\newcommand{\Pbf}{\mathbf{P}}
\newcommand{\Br}{\EuScript{B}}
\newcommand{\Dr}{\EuScript{D}}
\newcommand{\Fr}{\EuScript{F}}
\newcommand{\Mr}{\EuScript{M}}
\newcommand{\Nr}{\EuScript{N}}
\renewcommand{\Pr}{\EuScript{P}}
\newcommand{\dr}{\partial}
\newcommand{\Ptime}{\textsc{fptime}\xspace}

\definecolor{rouge}{rgb}{1,0,0}
\newcommand{\rouge}[1]{\textcolor{rouge}{#1}}
\definecolor{bleu}{rgb}{0,0,1}
\newcommand{\bleu}[1]{\textcolor{bleu}{#1}}

\newcommand{\figeps}[1]{\raisebox{-1.25mm}{\includegraphics{#1.eps}}}
\newcommand{\smallfigeps}[1]{\includegraphics[scale=0.66]{#1.eps}}

\def\doi{5 (2:14) 2009}
\lmcsheading%
{\doi}
{1--17}
{}
{}
{Jan.~\phantom{0}5, 2007}
{Jun.~\phantom{0}3, 2009}
{}   

\begin{document}

\title{Polygraphic programs and polynomial-time functions}

\author[G.~Bonfante]{Guillaume Bonfante}
\author[Y.~Guiraud]{Yves Guiraud}
\address{INRIA Nancy, 615 rue du Jardin Botanique, CS 20101, 54603 Villers-l\`es-Nancy, France}
\email{\{guillaume.bonfante,yves.guiraud\}@inria.fr}

\titlecomment{This work has been partially supported by ANR Inval project (ANR-05-BLAN-0267)}

\begin{abstract} We study the computational model of polygraphs. For
  that, we consider polygraphic programs, a subclass of these objects,
  as a formal description of first-order functional programs. We
  explain their semantics and prove that they form a Turing-complete
  computational model. Their algebraic structure is used by analysis
  tools, called polygraphic interpretations, for complexity
  analysis. In particular, we delineate a subclass of polygraphic
  programs that compute exactly the functions that are
  Turing-computable in polynomial time.
\end{abstract}

\maketitle

\section*{Introduction}

\subsection*{Polygraphs as a computational model} 
Polygraphs (or computads) are presentations by "generators" and
"relations" of some higher-dimensional
categories~\cite{Street76,Burroni93,Street87,Street95}. Albert Burroni
has proved that they provide an algebraic structure to equational
theories~\cite{Burroni93}. Yves Lafont and the second author have
explored some of the computational properties of these objects, mainly
termination, confluence and their links with term rewriting
systems~\cite{Lafont03,Guiraud06jpaa}. The present study, extending
notions and results presented earlier by the same
authors~\cite{BonfanteGuiraud07entcs}, concerns the complexity
analysis of polygraphs.

On a first approach, one can think of these objects as rewriting
systems on algebraic circuits: instead of computing on syntactical
terms, polygraphs make use of a net of cells, which individually
behave according to some local transition rules, as do John von
Neumann's cellular automata~\cite{vonNeumann66} and Yves Lafont's
interaction nets~\cite{Lafont90}.

Following Neil Jones' thesis that programming languages and semantics
have strong connexions with complexity theory~\cite{Jones97}, we think
that the syntactic features offered by polygraphs, with respect to
terms, play an important role from the point of view of implicit
computational complexity. As a running example, we consider the
divide-and-conquer algorithm of fusion sort. It computes the function
$f$ taking a list $l$ and returning the list made of the same
elements, yet sorted according to some given order relation. For that,
it uses a divide-and-conquer strategy: it splits $l$ into two sublists
$l_1$ and $l_2$ of equivalent sizes, then it recursively applies
itself on each one to get $f(l_1)$ and $f(l_2)$ and, finally, it
merges these two results to produce $f(l)$. The following program,
written in Caml~\cite{Caml}, implements this algorithm:
\medskip
\begin{verbatim}
  let rec split = function     
    | [] -> ([],[])
    | x::[] -> (x::[],[])
    | x::y::l -> let (l1,l2)=split(l) in (x::l1,y::l2)

  let rec merge = function    
    | ([],l) -> l
    | (l,[]) -> l
    | (x::l,y::m) -> if x<=y then x::merge(l,y::m) else y::merge(x::l,m) 

  let rec sort = function    
    | [] -> []
    | x::[] -> x::[] 
    | x::y::l -> let (l1,l2)=split(l) in merge(sort(x::l1),sort(y::l2))

\end{verbatim}

\noindent In a polygraph, one can consider, at the same level as other
operations, function symbols with many outputs. For example, the above
definition of the split function becomes, in the polygraphic language:
\begin{center}\input{tri-rapide-split.pstex_t}\end{center}

\noindent With these rules, one can actually "see" how the computation is made,
by "unzipping" lists. Also, one can internalize in polygraphs the
sharing operation of termgraphs~\cite{Plump99}, described as an
explicit and local duplication. As a consequence, the rules generating
computations become linear: the operations for pointers management can
be "seen" within the rules. Actually, in our analysis, we evaluate
explicitly the number of structural steps of computation: allocations,
deallocations and switches of pointers. In other words, we make
explicit the design of a garbage collector.

The question of sharing has been widely studied for efficient
implementations of functio\-nal programming languages and several
solutions have been suggested: for instance, Dan Dougherty, Pierre
Lescanne and Luigi Liquori proposed the formalism of addressed term
rewriting systems~\cite{DoughertyLescanneLiquori06}. Let us mention
another approach for this kind of issues due to Martin
Hofmann~\cite{Hofmann00}: he developed a typing discipline, with a
diamond type, for a functional language which allows a compilation
into an imperative language such as~C, without dynamic allocation.

The computational model of polygraphic programs, a subclass of
polygraphs, is explai\-ned in the first part of this document, where we
give their semantics and prove a completeness result: every
Turing-computable function can be computed by a polygraphic program.

\subsection*{Complexity analysis of polygraphic programs} 
Here we use tools inspired by polyno\-mial interpretations, which have
been introduced by Dallas Lankford to prove termination of term
rewriting systems~\cite{Lankford79}. They associate to each term a
polynomial with natural numbers as coefficients, in a way that is
naturally compatible with contexts and substitutions. When, for each
rule, the interpretation of the left-hand side is greater than the one
of the right-hand side, one gets a termination proof. For example, let
us consider the following term rewriting system that computes the
double function on natural numbers:
$$
d(0) \: \fl \: 0 \qquad\qquad d(s(x)) \: \fl \: s(s(d(x))).
$$
One proves its termination with the interpretation defined by
$\phi(0)=1$, $\phi(s(x))=\phi(x)+1$ and $\phi(d(x))=3\phi(x)$. Indeed, 
one checks that the following inequalities hold:
$$
\phi(d(0)) = 3 > 1 = \phi(0) 
\qquad\text{and}\qquad
\phi(d(s(x))) = 3\phi(x)+3 > 3\phi(x)+2 = \phi(s(s(d(x)))).
$$

\noindent Moreover, on top of termination results, polynomial
interpretations can be used to study complexity. For instance, Dieter
Hofbauer and Clemens Lautemann have established a doubly exponential
bound on the derivation length of systems with polynomial
interpreta\-tions \cite{HofbauerLautemann88}. Adam Cichon and Pierre
Lescanne have considered more precisely the compu\-ta\-tional power of
these systems~\cite{CichonLescanne92}. Adam Cichon, Jean-Yves Marion
and Hélène Touzet, with the first author, have identified complexity
classes by means of restrictions on polynomial
interpretations~\cite{BonfanteCichonMarionTouzet98,BonfanteCichonMarionTouzet01}.

Let us explain how this works on the example of the double
function. The given interpretation sends the term $d(s^n(0))$ to the
natural number $3n+3$: since each rule application will strictly
decrease this number, one knows that it takes at most $3n+3$ steps to
get from this term to its normal form $s^{2n}(0)$. Actually, the
considered interpretation gives a polynomial bound, with respect to
the size of the argument, on the time taken to compute the double
function with this program.

In order to analyze polygraphs, we use algebraic tools called
polygraphic interpretations, which have been introduced to prove
termination of polygraphs~\cite{Guiraud06jpaa}. Intuitively, one
considers that circuits are crossed by electrical currents. Depending
on the intensity of the currents that arrive to it, each circuit gate
produces some heat. Then one compares circuits according to the total
heat each one produces. Building a polygraphic interpretation amounts
at fixing how currents are transmitted by each gate and how much heat
each one emits.

The current part is called a functorial interpretation. Algebraically,
it is similar to a polynomial interpretation of terms and we also use
it as an estimation of the size of values, like
quasi-interpretations~\cite{BonfanteMarionMoyen05}. The heat part is
called a differential interpretation and it is specific to the
algebraic structure of polygraphs. We use it to bound the number of
computation steps remaining before reaching a result. Let us note that
the distinction between these two parts makes it possible for
polygraphic interpretations to cope with non-simplifying termination
proofs, like Thomas Arts and Jürgen Giesl's dependency
pairs \cite{ArtsGiesl00}.

However, some new difficulties arise with polygraphs. For example,
since duplication and erasure are explicit in our model, we must show
how to get rid of them for the interpretation. In our setting, the
programmer focuses on computational steps (as opposed to structural
steps) for which he has to give an interpretation. From this
interpretation, we give a polynomial upper bound on the number of
structural steps that will be performed.

In this work, we focus on polynomial-time computable functions or,
shorter, \Ptime functions. The reason comes from Stephen Cook's thesis
stating that this class corresponds to feasible computable
functions. But it is strongly conjectured that the preliminary results
developed in this paper can be used for other characterizations. In
particular, the current interpretations can be seen as
sup-interpretations, following~\cite{MarionPechoux06}: this means that
values have polynomial size.

Coming back to \Ptime, in the field of implicit computational
complexity, the notion of stratification has shown to be a fundamental
tool of the discipline. This has been developed by Daniel Leivant and
Jean-Yves Marion~\cite{Leivant91,LeivantMarion93} and by Stephen
Bellantoni and Stephen Cook~\cite{BellantoniCook92} to delineate
\Ptime. Other characterizations include Neil Jones' "Life without
cons" WHILE programs~\cite{Jones99} and Karl-Heinz Niggl and Henning
Wunderlich's characterization of imperative
programs~\cite{NigglWunderlich06}. There is also a logical approach to
implicit computational complexity, based on a linear type discipline,
in the seminal work of Jean-Yves Girard on light linear
logic~\cite{Girard98}, Yves Lafont on soft linear
logic~\cite{Lafont04} or Patrick Baillot and Kazushige
Terui~\cite{BaillotTerui04}.

The second part of this document is devoted to general results about
polygraphic interpretations of polygraphs. There, we explore the
pieces of information they can give us about size issues. Then, in the
third part, we apply these results to polygraphic programs. In
particular, we identify a subclass $\Pbf$ of these objects that
compute exactly the functions that can be computed in polynomial-time
by a Turing machine, or \Ptime functions for short.

\subsection*{General notations}
Throughout this document, we use several notations that we prefer to
group here for easier further reference.

If $X$ is a set and $p$ is a natural number, we denote by $X^p$ the cartesian product of $p$ copies of $X$. If $X$ is an ordered set, we equip $X^p$ with the product order, which is defined by $(x_1,\dots,x_p)\leq (y_1,\dots,y_p)$ whenever $x_i\leq y_i$ holds for every $i\in\ens{1,\dots,p}$.

If $f:X\fl X'$ and $g:Y\fl Y'$ are maps, then $f\times g$ denotes the
product map from $X\times X'$ to $Y\times Y'$. Let $f,g:X\fl Y$ be two
maps. If $Y$ is equipped with a binary relation $\vartriangleleft$,
then one compares $f$ and $g$ pointwise, which means that $f
\vartriangleleft g$ holds when, for every $x\in X$, one has
$f(x)\vartriangleleft g(x)$ in~$Y$. Similarly, if~$Y$ is equipped with
a binary operation $\diamond$, then one defines $f\diamond g$ as the
map from $X$ to $Y$ sending each~$x$ of~$X$ to the element
$f(x)\diamond g(x)$ in $Y$.

The sets $\Nb$ of natural numbers and $\Zb$ of integers are always
assumed to be equipped with their natural order. For every $n$ in
$\Nb$, we denote by $\mu_n$ the maximum map $\max\ens{x_1,\dots,x_n}$
and by $\Nb[x_1,\dots,x_n]$ the set of polynomials over $n$ variables
and with coefficients in $\Nb$. If $f:\Nb^m\fl\Nb^n$ is a map and if
$k\in\Nb$, one denotes by $kf$ the map sending $(x_1,\dots,x_m)$ to
$(ky_1,\dots,ky_n)$, if $(y_1,\dots,y_n)$ is $f(x_1,\dots,x_m)$.

\section{A computational model based on polygraphs}
\label{Section:Computation}

\subsection{A first glance at polygraphs}
\label{Subsection:IntroPoly}

On a first approach, one can consider polygraphs as rewriting systems on algebraic circuits, made of:

\subsection*{Types}
They are the wires, called $1$-cells. Each one conveys information of
some elementary type. To represent product types, one uses several
wires, in parallel, calling such a construc\-tion a $1$-path. For
example, the following $1$-path represents the type of quadruples made
of an integer, a boolean, a real number and a boolean:
$$
\rotatebox{90}{\raisebox{-1.25mm}{\begin{picture}(0,0)%
\includegraphics{1-cell-int.pstex}%
\end{picture}%
\setlength{\unitlength}{4144sp}%
\begingroup\makeatletter\ifx\SetFigFont\undefined%
\gdef\SetFigFont#1#2#3#4#5{%
  \reset@font\fontsize{#1}{#2pt}%
  \fontfamily{#3}\fontseries{#4}\fontshape{#5}%
  \selectfont}%
\fi\endgroup%
\begin{picture}(294,85)(124,737)
\put(271,762){\makebox(0,0)[b]{\smash{{\SetFigFont{6}{7.2}{\rmdefault}{\mddefault}{\updefault}{\color[rgb]{0,0,0}$\mathtt{int}$}%
}}}}
\end{picture}%
}}
\quad
\rotatebox{90}{\raisebox{-1.25mm}{\begin{picture}(0,0)%
\includegraphics{1-cell-bool.pstex}%
\end{picture}%
\setlength{\unitlength}{4144sp}%
\begingroup\makeatletter\ifx\SetFigFont\undefined%
\gdef\SetFigFont#1#2#3#4#5{%
  \reset@font\fontsize{#1}{#2pt}%
  \fontfamily{#3}\fontseries{#4}\fontshape{#5}%
  \selectfont}%
\fi\endgroup%
\begin{picture}(294,85)(124,737)
\put(271,762){\makebox(0,0)[b]{\smash{{\SetFigFont{6}{7.2}{\rmdefault}{\mddefault}{\updefault}{\color[rgb]{0,0,0}$\mathtt{bool}$}%
}}}}
\end{picture}%
}}
\quad
\rotatebox{90}{\raisebox{-1.25mm}{\begin{picture}(0,0)%
\includegraphics{1-cell-real.pstex}%
\end{picture}%
\setlength{\unitlength}{4144sp}%
\begingroup\makeatletter\ifx\SetFigFont\undefined%
\gdef\SetFigFont#1#2#3#4#5{%
  \reset@font\fontsize{#1}{#2pt}%
  \fontfamily{#3}\fontseries{#4}\fontshape{#5}%
  \selectfont}%
\fi\endgroup%
\begin{picture}(294,85)(124,737)
\put(271,762){\makebox(0,0)[b]{\smash{{\SetFigFont{6}{7.2}{\rmdefault}{\mddefault}{\updefault}{\color[rgb]{0,0,0}$\mathtt{real}$}%
}}}}
\end{picture}%
}}
\quad
\rotatebox{90}{\raisebox{-1.25mm}{\begin{picture}(0,0)%
\includegraphics{1-cell-bool.pstex}%
\end{picture}%
\setlength{\unitlength}{4144sp}%
\begingroup\makeatletter\ifx\SetFigFont\undefined%
\gdef\SetFigFont#1#2#3#4#5{%
  \reset@font\fontsize{#1}{#2pt}%
  \fontfamily{#3}\fontseries{#4}\fontshape{#5}%
  \selectfont}%
\fi\endgroup%
\begin{picture}(294,85)(124,737)
\put(271,762){\makebox(0,0)[b]{\smash{{\SetFigFont{6}{7.2}{\rmdefault}{\mddefault}{\updefault}{\color[rgb]{0,0,0}$\mathtt{bool}$}%
}}}}
\end{picture}%
}}
$$

\noindent The $1$-paths can be composed in one way, by putting them in parallel:
$$
\raisebox{-1.25mm}{\input{1-chemins-0-composition.pstex_t}}
$$

\subsection*{Operations}
They are represented by circuits, called $2$-paths. The gates used to
build them are called $2$-cells. The $2$-paths can be composed in two
ways, either by juxtaposition (parallel composition) or by connection
(sequential composition):
$$
\raisebox{-1.25mm}{\input{2-compositions.pstex_t}}
$$

\noindent Each $2$-path (or $2$-cell) has a finite number of typed
inputs, a $1$-path called its $1$-source, and a finite number of typed
outputs, a $1$-path called its $1$-target:
$$
\raisebox{-1.25mm}{\input{2-cell-1-bord.pstex_t}}
$$

\noindent Several constructions represent the same operation. In
particular, wires can be stretched or contracted, provided one does
not cross them or break them. This can be written either graphically
or algebraically:
$$
\raisebox{-1.25mm}{\input{2-cellules-01-echange.pstex_t}}
$$
$$
\big( f \star_0 s_1(g) \big) \star_1 \big( t_1(f) \star_0 g \big) 
\quad\equiv\quad
f\star_0 g
\quad\equiv\quad
\big( s_1(f) \star_0 g \big) \star_1 \big( f \star_0 t_1(g) \big).
$$

\subsection*{Computations}
They are rewriting paths, called $3$-paths, transforming a given
$2$-path, called its $2$-source, into another one, called its
$2$-target. The $3$-paths are generated by local rewriting rules,
called $3$-cells. The $2$-source and the $2$-target of a $3$-cell or
$3$-path are required to have the same input and output, \ie the same
$1$-source and the same $1$-target. A $3$-path is represented either
as a reduction on $2$-paths or as a genuine $3$-dimensional object:
$$
\raisebox{-1.25mm}{\input{coupes_bloc.pstex_t}}
\qquad\qquad\qquad\qquad
\raisebox{-3mm}{\raisebox{-1.25mm}{\input{bloc.pstex_t}}}
$$
The $3$-paths can be composed in three ways, two parallel
ones coming from the structure of the $2$-paths, plus one new,
sequential one:
\begin{center}
$\raisebox{-1.25mm}{\input{3-chemins-0-composition.pstex_t}}$ \\
$\raisebox{-1.25mm}{\input{3-chemins-1-composition.pstex_t}} 
\qquad\qquad\qquad
\raisebox{-1.25mm}{\input{3-chemins-2-composition.pstex_t}}$
\end{center}
The $3$-paths are identified modulo relations that include topological
moves such as:
$$
\raisebox{-1.25mm}{\input{echanges_blocs.pstex_t}}
$$

\noindent These graphical relations have an algebraic version given, for $0\leq i<j\leq 2$, by:
$$
\big( F \star_i s_j(G) \big) \star_j \big( t_j(F) \star_i G \big) 
\quad\equiv\quad
F \star_i G
\quad\equiv\quad
\big( s_j(F) \star_i G \big) \star_j \big( F \star_i t_j(G) \big).
$$

\noindent So far, we have described a special case of
$3$-polygraphs. A $n$-polygraph is a similar object, made of cells,
paths, sources, targets and compositions in all dimensions up to $n$.

\begin{rem}
Polygraphs provide a uniform, algebraic and graphical description of
objects coming from different domains: abstract, string and term
rewriting systems~\cite{Lafont03,Guiraud04,Guiraud06jpaa}, abstract
algebraic structures~\cite{Burroni93,Guiraud04,Loday06}, Feynman and
Penrose dia\-grams \cite{BaezLauda06}, braids, knots and tangle
diagrams equipped with Reidemeister
moves \cite{Adams04,Guiraud04}, Petri nets~\cite{Guiraud06tcs} and
propositional proofs of classical and linear
logics~\cite{Guiraud06apal}.
\end{rem}

\subsection{Polygraphs}
\label{Subsection:Polygraphs}

On a first reading, one can skip the formal definition of polygraph
and just keep in mind the graphical introduction. We define
$n$-polygraphs by induction on the dimension~$n$: given a definition
of $(n-1)$-polygraphs, we define a $n$-polygraph as a base
$(n-1)$-polygraph extended with a set of $n$-cells. Let us initiate
the induction with $0$-polygraphs and $1$-polygraphs.

\begin{defi} \label{Definition:ZeroPolygraph}
A \emph{$\mathit{0}$-polygraph} is a set $\Pr$. Its
\emph{$\mathit{0}$-cells} and \emph{$\mathit{0}$-paths} are its
elements.
\end{defi}

\begin{defi} \label{Definition:OnePolygraph}
A \emph{$\mathit{1}$-polygraph} is a data $\Pr=(\Br,\Pr_1,s,t)$ made
of a $0$-polygraph $\Br$, a set~$\Pr_1$ and two maps $s$ and $t$ from
$\Pr_1$ to $\Br$. The \emph{$\mathit{0}$-cells} and
\emph{$\mathit{0}$-paths} of $\Pr$ are the ones of $\Br$. Its
\emph{$\mathit{1}$-cells} are the elements of $\Pr_1$. One inductively
defines the set $\mon{\Pr_1}$ of \emph{$\mathit{1}$-paths} of $\Pr$,
together with the \emph{$\mathit{0}$-source map} $s_0$ and the
\emph{$\mathit{0}$-target map} $t_0$, both from $1$-paths to
$0$-paths, as follows:
\begin{enumerate}[$\bullet$]
\item Every $0$-cell $x$ is a $1$-path, with $s_0(x)=t_0(x)=x$.
\item Every $1$-cell $\xi$ is a $1$-path, with $s_0(\xi)=s(\xi)$ and $t_0(\xi)=t(\xi)$.
\item If $u$ and $v$ are $1$-paths such that $t_0(u)=s_0(v)$, then $u\star_0 v$ is a $1$-path called the \emph{$\mathit{0}$-composition} of~$u$ and $v$. One defines $s_0(u\star_0v)=s_0(u)$ and $t_0(u\star_0v)=t_0(v)$. 
\end{enumerate}

\noindent The $1$-paths are identified modulo the following relations:
\begin{enumerate}[$\bullet$]
\item Associativity: $\:(u\star_0 v)\star_0 w=u\star_0(v\star_0 w)\:$.
\item Local units: $\:s_0(u)\star_0 u=u=u\star_0 t_0(u) \:$.
\end{enumerate}
\end{defi}

\begin{exa}
A graph yields a $1$-polygraph, with vertices as $0$-cells and arrows as $1$-cells. The $1$-paths are the paths in the graph. 
\end{exa}

\begin{exa}
A set $X$ can be seen as a $1$-polygraph, with one $0$-cell and itself as set of $1$-cells: in that case, the set $\mon{X}$ of $1$-paths is exactly the free monoid generated by~$X$ or, equivalently, the set of words over the alphabet $X$.
\end{exa}

\begin{exa}
An abstract rewriting system is a binary relation $R$ over a set $X$. Such an object yields a $1$-polygraph $\Pr$ with $\Pr_0=X$, $\Pr_1=R$, $s_0(x,y)=x$ and $t_0(x,y)=y$. Then, the $1$-paths of this $1$-polygraph are in bijective correspondence with the rewriting paths generated by $(X,R)$.
\end{exa}

\noindent Now, let us fix a natural number $n\geq 2$ and assume that one has
defined what a $(n-1)$-polygraph~$\Pr$ is, how one builds its sets
$\Pr_k$ of $k$-cells and $\mon{\Pr_k}$ of $k$-paths,
$k\in\ens{0,\dots,n}$, and its $j$-source map~$s_j$ and $j$-target
map~$t_j$ from $\mon{\Pr_k}$ to $\mon{\Pr_j}$,
$j\in\ens{0,\dots,k-1}$.

\begin{defi} \label{Definition:Polygraph}
  An \emph{$\mathit{n}$-polygraph} is a data $\Pr=(\Br,\Pr_n,s,t)$
  made of an $(n-1)$-polygraph~$\Br$, a set~$\Pr_n$ and two maps $s$
  and $t$ from $\Pr_n$ to $\mon{\Br_{n-1}}$, such that the
  \emph{globular relations} hold:
\[s_{n-2} \circ s = s_{n-2} \circ t 
  \qquad \text{and} \qquad
  t_{n-2} \circ s = t_{n-2} \circ t.
\]
  For every $k$ in $\ens{0,\dots,n-1}$, the \emph{$\mathit{k}$-cells}
  and \emph{$\mathit{k}$-paths} of $\Pr$ are the ones of $\Br$. The
  \emph{$\mathit{n}$-cells} of~$\Pr$ are the elements of $\Pr_n$. One
  inductively defines the set $\mon{\Pr_n}$ of
  \emph{$\mathit{n}$-paths} of $\Pr$, the
  \emph{$\mathit{(n-1)}$-source map}~$s_{n-1}$, the
  \emph{$\mathit{(n-1)}$-target map}~$t_{n-1}$ and, for every
  $k\in\ens{0,\dots,n-2}$, extensions to $n$-paths of the $k$-source
  map~$s_k$ and the $k$-target map~$t_k$ of~$\Br$:

\begin{enumerate}[$\bullet$]
\item For every $k\in\ens{0,\dots,n-1}$, every $k$-cell $\xi$ is an
  $n$-path, with $s_{n-1}(\xi)=t_{n-1}(\xi)=\xi$. Values of other
  source and target maps do not change.

\item Every $n$-cell $\phi$ is an $n$-path, with
  $s_{n-1}(\phi)=s(\phi)$ and $t_{n-1}(\phi)=t(\phi)$. If
  $k\in\ens{0,\dots,n-2}$, then $s_k$ and $t_k$ are respectively
  extended by $s_k(\phi)=s_k\circ s_{n-1}(\phi)$ and by
  $t_k(\phi)=t_k\circ t_{n-1}(\phi)$.

\item If $k\in\ens{0,\dots,n-1}$ and if $f$ and $g$ are $n$-paths such that $t_k(f)=s_k(g)$ holds, then $f\star_k g$ is an $n$-path called the \emph{$\mathit{k}$-composition} of $f$ and $g$. For $j\in\ens{0,\dots,n-2}$, one defines:
$$
s_j(f\star_k g) = 
\begin{cases}
s_j(f) &\text{if }j\leq k \\
s_j(f)\star_k s_j(g) &\text{if }j>k
\end{cases}
\qquad\text{and}\qquad
t_j(f\star_k g) = 
\begin{cases}
t_j(g) &\text{if }j\leq k \\
t_j(f)\star_k t_j(g) &\text{if }j>k.
\end{cases}
$$
\end{enumerate}

\noindent One does not distinguish two $n$-paths that only differ by the following relations:
\begin{enumerate}[$\bullet$]
\item Associativity: $\:(f\star_k g)\star_k h = f\star_k (g\star_k h),\:$ for $0\leq k\leq n-1$.
\item Local units: $\:s_k(f) \star_k f = f = f\star_k t_k(f),\:$ for $0\leq k\leq n-1$.
\item Exchange: $\:(f_1\star_j f_2)\star_k(g_1\star_j g_2)=(f_1\star_k g_1)\star_j(f_2\star_k g_2),\:$ for $0\leq j<k\leq n-1$.
\end{enumerate}
\end{defi}

\begin{exa}
Let us consider a word rewriting system $(X,R)$, made of set $X$ and a
binary relation~$R$ over~$\mon{X}$. From it, one builds a
$2$-polygraph $\Pr$ with one $0$-cell, $\Pr_1=X$, $\Pr_2=R$,
$s_1(u,v)=u$ and $t_1(u,v)=v$. There is a bijection between the
$2$-paths of $\Pr$ and the rewriting paths generated by $(X,R)$,
considered modulo the commutation squares between two non-overlapping
rule applications. Moreover the circuit-like pictures provide
graphical representations for word rewriting: wires are letters, gates
are applications of rewriting rules and circuits are traces of
computations.
\end{exa}

\begin{exa}
Term rewriting systems generate $3$-polygraphs, as explained by Albert
Burroni~\cite{Burroni93}, Yves Lafont~\cite{Lafont03} and the second
author~\cite{Guiraud06jpaa,Guiraud06apal}. The poly\-gra\-phic programs
one considers here are light versions of these~\cite{Guiraud07}.
\end{exa}

\begin{exa}
Petri nets correspond exactly to $3$-polygraphs with one $0$-cell and
no $1$-cell: one identifies places with $2$-cells and transitions with
$3$-cells~\cite{Guiraud06tcs}.
\end{exa}

\begin{defi}
Let us fix a natural number $n$ and an $n$-polygraph $\Pr$. The
polygraph $\Pr$ is \emph{finite} when it has a finite number of cells
in every dimension. A family $X$ of $n$-cells of $\Pr$ can be seen as
an $n$-polygraph with the same cells as $\Pr$ up to dimension $n-1$.

If $0\leq j<k\leq n$, two $k$-paths~$f$ and~$g$ are
\emph{$\mathit{j}$-composable} when $t_j(f)=s_j(g)$. They are
\emph{$\mathit{j}$-parallel} when $s_j(f)=s_j(g)$ and
$t_j(f)=t_j(g)$. When $j=k-1$, one simply says \emph{composable} and
\emph{parallel}. Similarly, the $(k-1)$-source and $(k-1)$-target of a
$k$-path are simply called its \emph{source} and \emph{target}.

If $0\leq k\leq n$, given a subset $X$ of $\Pr_k$ and a $k$-path $f$,
the \emph{size of $\mathit{f}$ with respect to $\mathit{X}$} is the
natural number denoted by~$\norm{f}_X$ and defined as follows, by
structural induction on~$f$:
$$
\norm{f}_X \:=\:
\begin{cases}
0 &\text{if $f$ is a cell and $f\notin X$,} \\
1 &\text{if $f\in X$,} \\
\norm{g}_X+\norm{h}_X &\text{if $f=g\star_j h$, for some $0\leq j<k$.}
\end{cases}
$$

\noindent When $X$ is reduced to one cell $\phi$, one writes
$\norm{f}_{\phi}$ instead of $\norm{f}_{\ens{\phi}}$. The \emph{size
  of $\mathit{f}$} is its size with respect to~$\Pr_k$, simply written
$\norm{f}$. A $k$-path is \emph{degenerate} when it has size~$0$ and
\emph{elementary} when its size is~$1$.
\end{defi}

\begin{rem}
One must check that the definition of the size of a $k$-path (with
respect to a set of $k$-cells $X$) is correct. This is done by
computing this map on both sides of the relations of associativity,
local units and exchange and ensuring that both results are equal.

One proves that any non-degenerate $k$-path $f$ of size $p$ can be written 
$$
f \:=\: f_1 \star_{k-1} \cdots \star_{k-1} f_p,
$$

\noindent where each~$f_i$ is an elementary $k$-path. Moreover, if
$k\geq 1$, then any elementary $k$-path~$f$ can be written as follows:
$$
f \:=\: g_k \star_{k-1} \big( g_{k-1} \star_{k-2} \cdots \star_1 ( g_1\star_0 \phi \star_0 h_1) \star_1 \cdots \star_{k-2} h_{k-1} \big) \star_{k-1} h_k,
$$
where $\phi$ is a uniquely defined $k$-cell, while $g_j$ and
$h_j$ are $j$-paths, for every $j\in\ens{1,\dots,k}$. For example, any
elementary $3$-path $F$ can be decomposed as $F=f\star_1(u\star_0
\alpha \star_0 v)\star_1 g$, where~$\alpha$ is a uniquely determined
$3$-cell, $f$ and $g$ are $2$-paths, $u$ and $v$ are $1$-paths. As a
consequence:
$$
\raisebox{-1.25mm}{\input{decomposition-reduction-f.pstex_t}} \qquad\qquad\qquad\qquad 
\raisebox{-1.25mm}{\input{decomposition-reduction-g.pstex_t}}
$$
$$
s_2 F \:=\: f\star_1(u\star_0 s_2 \alpha \star_0 v)\star_1 g \qquad\qquad\qquad 
t_2 F \:=\: f\star_1(u\star_0 t_2 \alpha \star_0 v)\star_1 g
$$
\end{rem}

\noindent In order to study the computational properties of
polygraphs, we use notions of higher-di\-men\-sional rewriting
theory~\cite{Guiraud06jpaa} that, in turn, make reference to abstract
rewriting ones~\cite{BaaderNipkow98}.

\begin{defi}
The \emph{reduction graph} associated to an $n$-polygraph $\Pr$ is the
graph with $(n-1)$-paths of $\Pr$ as objects and elementary $n$-paths
of $\Pr$ as arrows. Rewriting notions of \emph{normal forms},
\emph{termination}, \emph{(local) confluence}, \emph{convergence},
etc. are defined on $\Pr$ by taking back the ones of its reduction
graph.
\end{defi}

\begin{rem}
One can check that, given two parallel $(n-1)$-paths $f$ and $g$ in an
$n$-poly\-graph $\Pr$, there exists a path from $f$ to $g$ in the
reduction graph of $\Pr$ if and only if there exists a non-degenerate
$n$-path $F$ with source $f$ and target $g$ in $\Pr$.
\end{rem}

\noindent In what follows, we focus on $3$-polygraphs and introduce some special
notions and notations for them.

\begin{defi}
Let $\Pr$ be a $3$-polygraph. The fact that $f$ is a $k$-path of $\Pr$ with source $x$ and target $y$ is denoted by $f:x\fl y$ when $k=1$, by $f:x\dfl y$ when $k=2$, by $f:x\tfl y$ when $k=3$. If~$f$ is a $k$-path of $\Pr$ and $X$ a family of $k$-cells then, instead of $\norm{f}_X$, one writes~$\abs{f}_X$ when $k=1$ and $\tnorm{f}_X$ when $k=3$. When~$f:x\dfl y$, then $\abs{x}$, $\abs{y}$ and $(\abs{x},\abs{y})$ are respectively called the \emph{arity}, the \emph{coarity} and the \emph{valence} of $f$. 
\end{defi}

\subsection{Polygraphic programs}
\label{Subsection:Programs}

\begin{defi}\label{Definition:PolygraphicProgram}
A \emph{polygraphic program} is a finite $3$-polygraph $\Pr$ with one $0$-cell, thereafter denoted by~$\ast$, and such that its sets of $2$-cells and of $3$-cells respectively decompose into $\Pr_2=\Pr_2^S\amalg\Pr_2^C\amalg\Pr_2^F$ and $\Pr_3=\Pr_3^S\amalg\Pr_3^R$, with the following conditions:

\begin{enumerate}[$\bullet$]

\item The set $\Pr_2^S$ is made of the following elements, called \emph{structure $\mathit{2}$-cells}, where $\xi$ and~$\zeta$ range over the set of $1$-cells of $\Pr$:
$$
\figeps{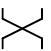}_{\xi,\zeta}:\xi\star_0\zeta \dfl \zeta\star_0\xi,
\qquad
\figeps{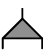}_{\xi}:\xi \dfl \xi\star_0\xi,
\qquad
\figeps{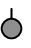}_{\xi}:\xi \dfl \ast.
$$ 

\noindent When the context is clear, one simply writes $\figeps{tau}$, $\figeps{delta}$ and $\figeps{epsilon}$. The following elements of~$\mon{\Pr_2^S}$ are called \emph{structure $2$-paths} and they are defined by structural induction on their $1$-source:
\begin{center}\input{2-chemins-structure.pstex_t}\end{center}

\smallskip
\item The set $\Pr_2^C$ is made of $2$-cells with coarity $1$, \ie of the shape $\figeps{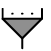}$, called \emph{constructor $\mathit{2}$-cells}. 

\item The elements of $\Pr_2^F$ are called \emph{function $\mathit{2}$-cells}.

\item The elements of $\Pr_3^S$, called \emph{structure $\mathit{3}$-cells}, are defined, for every constructor $2$-cell $\figeps{phi}:x\dfl\xi$ and every $1$-cell $\zeta$, by:
\begin{center}\input{3-cellules-structure-2.pstex_t}\end{center}

\item The elements of $\Pr_3^R$ are called \emph{computation $\mathit{3}$-cells} and each one has a $2$-source of the shape $t\star_1\figeps{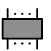}$, with $t\in\mon{\Pr_2^C}$ and $\figeps{fonction}\in\Pr_2^F$.
\end{enumerate}
\end{defi}

\begin{rem}
In this study, we have decided to split structure cells from computation cells. From a traditional programming perspective, permutations, duplications and erasers are given for free in the syntax. With polygraphs, this is not the case. However, by putting these operations in a "special" sublayer, we show that the programmer has not to bother with structure cells: one can stay at the top-level, letting the sublevel work on its own. 
\end{rem}

\begin{exa}\label{Example:Division}
The following polygraphic program $\Dr$ computes the euclidean division on natural numbers (we formally define what this means later):
\begin{enumerate}[(1)]
\item It has one $1$-cell $\mathtt{n}$, standing for the type of natural numbers.
\item Apart from the fixed three structure $2$-cells, it has two constructor $2$-cells, $\figeps{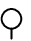}:\ast\dfl\mathtt{n}$ for zero and $\figeps{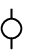}:\mathtt{n}\dfl\mathtt{n}$ for the successor operation, and two function $2$-cells,  $\figeps{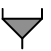}:\mathtt{n}\star_0\mathtt{n}\dfl\mathtt{n}$ for the minus function and $\figeps{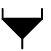}:\mathtt{n}\star_0\mathtt{n}\dfl\mathtt{n}$ for the division function.
\item Its $3$-cells are made of eight structure $3$-cells, plus the following five computation $3$-cells:
\begin{center}\input{3-cellules-quotient.pstex_t}\end{center}
\end{enumerate}
\end{exa}

\begin{exa}\label{Example:FusionSort}
The following program $\Fr$ computes the \emph{fusion sort} function on lists of natural numbers lower or equal than some constant $N\in\Nb$: 

\begin{enumerate}[(1)]

\item Its $1$-cells are $\mathtt{n}$, for natural numbers, and $\mathtt{l}$, for lists of natural numbers. 

\item Its $2$-cells are made of eight structure $2$-cells, plus:

\begin{enumerate}[(a)]
\item Constructor $2$-cells, for the natural numbers $0$, $\dots$, $N$, the empty list and the list constructor:
$$
\left(\raisebox{-1.25mm}{\begin{picture}(0,0)%
\includegraphics{n.pstex}%
\end{picture}%
\setlength{\unitlength}{4144sp}%
\begingroup\makeatletter\ifx\SetFigFont\undefined%
\gdef\SetFigFont#1#2#3#4#5{%
  \reset@font\fontsize{#1}{#2pt}%
  \fontfamily{#3}\fontseries{#4}\fontshape{#5}%
  \selectfont}%
\fi\endgroup%
\begin{picture}(139,174)(887,1097)
\put(957,1188){\makebox(0,0)[b]{\smash{{\SetFigFont{6}{7.2}{\rmdefault}{\mddefault}{\updefault}{\color[rgb]{0,0,0}$n$}%
}}}}
\end{picture}%
}:\ast\dfl\mathtt{n}\right)_{0\leq n\leq N}, \qquad
\figeps{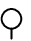}:\ast\dfl\mathtt{l}, \qquad
\figeps{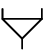}:\mathtt{n}\star_0\mathtt{l}\dfl\mathtt{l}.
$$

\item Function $2$-cells, respectively for the main sort and the two auxiliary split and merge:
$$
\figeps{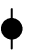}:\mathtt{l}\dfl\mathtt{l}, \qquad
\figeps{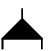}:\mathtt{l}\dfl\mathtt{l}\ast_0\mathtt{l}, \qquad
\figeps{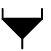}:\mathtt{l}\ast_0\mathtt{l}\dfl\mathtt{l}.
$$
\end{enumerate}

\item Its $3$-cells are made of $6N+18$ structure $3$-cells, plus $N^2+2N+8$ computation $3$-cells:
\begin{center}\input{tri-rapide-sort.pstex_t}\end{center}
\begin{center}\input{tri-rapide-split.pstex_t}\end{center}
\begin{center}\input{tri-rapide-merge.pstex_t}\end{center}
\end{enumerate}
\end{exa}

\begin{rem}
One may object that sorting lists when the a priori bound~$N$ is known can be performed in a linear number of steps: one reads the list and counts the number of occurrences of each element, then produces the sorted list from this information. Neverthe\-less, the presented algorithm (up to the test $\leq$ on the natural numbers $p$ and $q$) really mimics the "mechanics" of the fusion sort algorithm and, actually, we rediscover the complexity bound as given by Yiannis Moschovakis~\cite{Moschovakis01}.

Why don't we internalize the comparison of numbers within the polygraphic program? This comes from the fact that the \emph{if-then-else} construction implicitly involves an evaluation strategy: one first computes the test argument then, depending on this result, one computes \emph{exactly one} of the other two arguments. As defined here, polygraphs algebraically describe the computation steps, but not the evaluation strategy. We let such a task for further research.
\end{rem}

\subsection{Semantics of polygraphic programs}
\label{Subsection:Semantics}

One defines an interpretation $\sem{\cdot}$ of the ele\-ments of a polygraphic program into sets and maps, then one uses it to define the notion of function computed by such a program.

\begin{defi}\label{Definition:Semantics}
Let $\Pr$ be a polygraphic program. For a $1$-path~$u$, a \emph{value
  of type~$\mathit{u}$} is a $2$-path in~$\mon{\Pr_2^C}$ with
source~$\ast$ and target~$u$; their set is denoted by~$\sem{u}$. Given
a $2$-path $f:u\dfl v$, one denotes by~$\sem{f}$ the (partial) map
from~$\sem{u}$ to~$\sem{v}$ defined as follows: if~$t$ is a value of
type~$u$ and if $t\star_1 f$ has a unique normal form~$t'$ that is a
value (of type $v$), then~$\sem{f}(t)$ is~$t'$; otherwise~$f$ is
undefined on~$t$.
\end{defi}

\noindent Among the following properties, the one for degenerate
$2$-paths explains the fact that~$\sem{u}$ has two meanings: it is
either the set of values of type $u$ or the identity of this set.

\begin{prop}\label{Proposition:Semantics}
Let $\Pr$ be a polygraphic program. The following properties hold on $1$-paths:
\begin{enumerate}[$\bullet$]
\item The set $\sem{\ast}$ is reduced to the $0$-cell $\ast$.
\item For every $u$ and $v$, one has $\sem{u\star_0 v}=\sem{u}\times\sem{v}$.
\end{enumerate}
The following properties hold on $2$-paths:
\begin{enumerate}[$\bullet$]
\item If $u$ is degenerate then it is sent by $\sem{\cdot}$ to the
  identity of the set~$\sem{u}$.
\item For every $f$ and $g$, one has $\sem{f\star_0 g}=\sem{f}\times\sem{g}$.
\item If $f$ and $g$ are composable, then $\sem{f\star_1
  g}=\sem{g}\circ\sem{f}$ holds.
\end{enumerate}
Finally, for every $3$-path $F$, the equality $\sem{s_2 F }=\sem{t_2 F}$ holds. 
\qed
\end{prop}

\begin{defi}
Let $\Pr$ be a polygraphic program. Let $u$, $v$ be $1$-paths and
let~$f$ be a (partial) map from~$\sem{u}$ to $\sem{v}$. One says
that~$\Pr$ \emph{computes~$f$} when there exists a $2$-cell
$\figeps{fonction}$ such that $\sem{\figeps{fonction}}=f$.
\end{defi}

\begin{exa}
In a polygraphic program $\Pr$, every constructor $2$-cell
$\figeps{phi}$ with arity $n$ satisfies the equality
$\sem{\figeps{phi}}(t_1,\dots,t_n)=(t_1\star_0\dots\star_0
t_n)\star_1\figeps{phi}$. Since the right member is always a normal
form, one can identify values of coarity~$1$ with the closed terms of
a term algebra. Moreover, the polygraphic program~$\Pr$ computes
erasers, duplications and permutations on these terms, since
$\sem{\figeps{epsilon}}(t)=\ast$, $\sem{\figeps{delta}}(t)=(t,t)$ and
$\sem{\figeps{tau}}(t,t')=(t',t)$ hold.
\end{exa}

\noindent Thus, every polygraphic program computes one total map for each of its structure and constructor $2$-cells. We give sufficient conditions to ensure that this is also the case on function $2$-cells.

\begin{defi}
A polygraphic program $\Pr$ is \emph{complete} if every $2$-path of the form $t\star_1\figeps{fonction}$ is reducible when $t$ is a value and $\figeps{fonction}$ is a function $2$-cell. 
\end{defi}

\begin{prop}\label{Proposition:FunctionsSemantics}
Let $\Pr$ be a convergent and complete polygraphic program. Then, for every structure or function $2$-cell $\figeps{fonction}:u\dfl v$, the map $\sem{\figeps{fonction}}:\sem{u}\fl\sem{v}$ is total.
\end{prop}

\proof We start by recalling that the structure $3$-cells, alone, are
convergent\,\cite{Guiraud06jpaa,Guiraud06apal}. Furthermore, they are
orthogonal to the computation $3$-cells and every $2$-path of the
shape $t\star_1\figeps{fonction}$ is reducible when $t$ is a value and
$\figeps{fonction}$ is a structure $2$-cell. Hence, as a
polygraph,~$\Pr$ is convergent and the $2$-paths $\ast\dfl x$ that are
in normal form are exactly the values of type~$x$.  \qed

\begin{exa}
Let us check that the polygraphic program $\Dr$ computes euclidean
division. The set $\sem{\mathtt{n}}$ is equipotent to the set $\Nb$ of
natural numbers through the bijection $\ul{0} \:=\: \figeps{nil}$ and
$\ul{n+1} \:=\: \ul{n}\star_1\figeps{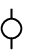}$. This polygraphic program
is weakly orthogonal, hence locally confluent, and complete. We will
also see later that it terminates. Thus it computes two maps from
$\sem{\mathtt{n}\star_0\mathtt{n}}\simeq\Nb^2$ to
$\sem{\mathtt{n}}\simeq\Nb$, one for~$\figeps{fonction-2-b}$ and one
for~$\figeps{fonction-2}$. By induction on the arguments, one gets:
$$
\sem{\figeps{fonction-2-b}} \left(\ul{m},\ul{n}\right)=\ul{\max\ens{0,m-n}}
\qquad\text{and}\qquad 
\sem{\figeps{fonction-2}} \left(\ul{m},\ul{n} \right)=\ul{\rounddown{m/(n+1)}}.
$$
\end{exa}

\begin{exa}\label{Example:Reduction}
\newcommand{\un}{\scriptscriptstyle 1}
\newcommand{\deux}{\scriptscriptstyle 2}
In the polygraphic program $\Fr$, one has $\sem{\mathtt{n}}\simeq\ens{0,\dots,N}$ and $\sem{\mathtt{l}}\simeq\mon{0,\dots,N}$, thanks to the bijective correspondences $\ul{n}=\raisebox{-1.25mm}{\begin{picture}(0,0)%
\includegraphics{n.pstex}%
\end{picture}%
\setlength{\unitlength}{4144sp}%
\begingroup\makeatletter\ifx\SetFigFont\undefined%
\gdef\SetFigFont#1#2#3#4#5{%
  \reset@font\fontsize{#1}{#2pt}%
  \fontfamily{#3}\fontseries{#4}\fontshape{#5}%
  \selectfont}%
\fi\endgroup%
\begin{picture}(139,174)(887,1097)
\put(957,1188){\makebox(0,0)[b]{\smash{{\SetFigFont{6}{7.2}{\rmdefault}{\mddefault}{\updefault}{\color[rgb]{0,0,0}$n$}%
}}}}
\end{picture}%
}$, $\ul{[\:]}=\figeps{nil}$ and $\ul{x::l}=(\ul{x}\star_0\ul{l})\star_1\figeps{cons}$. This polygraphic program is weakly orthogonal, hence locally confluent, and complete. It is also terminating, as we shall see later. Thus, it computes one map for each of~$\figeps{sort}$,~$\figeps{split}$ and~$\figeps{merge}$. For example, the map~$\sem{\figeps{sort}}$ takes a list of natural numbers as input and returns the corresponding ordered list. Figure~\ref{Figure:Computation} gives an example of computation generated by this program, with explanations following.
\begin{figure}[!htp]
\begin{align*}
\left( \raisebox{-1.25mm}{\input{2-1-nil.pstex_t}} \right) \star_1 \rouge{\boxed{\figeps{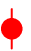}_3}}
\quad : &\quad
\raisebox{-4mm}{\raisebox{-1.25mm}{\input{ex-tri-1.pstex_t}}} 
\quad\tfl\quad 
\raisebox{-5mm}{\raisebox{-1.25mm}{\input{ex-tri-1b.pstex_t}}} \\ 
\star_2 \qquad & \\
\rouge{\boxed{\figeps{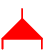}_1}} \star_1 \left( \raisebox{-1.25mm}{\input{2.pstex_t}} \star_0 \bleu{\boxed{\figeps{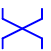}_{\:\scalebox{0.66}{\input{1-bleu.pstex_t}},\:\smallfigeps{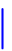}}}} \star_0 \figeps{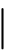} \right) \star_1 \raisebox{-3mm}{\figeps{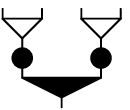}}
\quad : &\quad
\raisebox{-5mm}{\raisebox{-1.25mm}{\input{ex-tri-2.pstex_t}}}
\quad\tfl\quad 
\raisebox{-4mm}{\raisebox{-1.25mm}{\input{ex-tri-2b.pstex_t}}} \\ 
\star_2 \qquad & \\
\left( \raisebox{-1.25mm}{\input{2-1.pstex_t}} \right) \star_1 \left( \: \rouge{\boxed{\figeps{sort-rouge}_2}} \star_0 \bleu{\boxed{\figeps{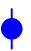}_2}} \: \right) \star_1 \figeps{merge} 
\quad : &\quad
\raisebox{-4mm}{\raisebox{-1.25mm}{\input{ex-tri-3.pstex_t}}}
\quad\tfl\quad 
\raisebox{-3mm}{\raisebox{-1.25mm}{\input{ex-tri-3b.pstex_t}}} \\
\star_2 \qquad & \\
\left( \figeps{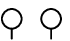} \right) \star_1 \rouge{\boxed{\figeps{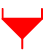}_3(\raisebox{-1.25mm}{\input{2-rouge.pstex_t}},\raisebox{-1.25mm}{\input{1-rouge.pstex_t}})}}
\quad : &\quad
\raisebox{-3mm}{\raisebox{-1.25mm}{\input{ex-tri-4.pstex_t}}} 
\quad\tfl\quad 
\raisebox{-4mm}{\raisebox{-1.25mm}{\input{ex-tri-4b.pstex_t}}} \\
\star_2 \qquad & \\
\raisebox{-1.5mm}{\raisebox{-1.25mm}{\input{ex-5-bout.pstex_t}}} \star_1 \left( \raisebox{-1.25mm}{\input{1.pstex_t}} \star_0 \rouge{\boxed{\figeps{merge-rouge}_2}} \: \right) \star_1 \figeps{cons}
\quad : &\quad
\raisebox{-4mm}{\raisebox{-1.25mm}{\input{ex-tri-5.pstex_t}}} 
\quad\tfl\quad 
\raisebox{-3mm}{\raisebox{-1.25mm}{\input{ex-tri-5b.pstex_t}}}.
\end{align*}
\caption{Normalizing $3$-path in a polygraphic program}
\label{Figure:Computation}
\end{figure}

\noindent Let us consider the list $[2;1]$ of natural numbers and apply the fusion sort function $\figeps{sort}$ on it. The list is coded by the following value:
$$
\ul{[2;1]} 
\:=\: 
\left( \raisebox{-1.25mm}{\begin{picture}(0,0)%
\includegraphics{1.pstex}%
\end{picture}%
\setlength{\unitlength}{4144sp}%
\begingroup\makeatletter\ifx\SetFigFont\undefined%
\gdef\SetFigFont#1#2#3#4#5{%
  \reset@font\fontsize{#1}{#2pt}%
  \fontfamily{#3}\fontseries{#4}\fontshape{#5}%
  \selectfont}%
\fi\endgroup%
\begin{picture}(106,204)(1118,1097)
\put(1171,1177){\makebox(0,0)[b]{\smash{{\SetFigFont{6}{7.2}{\rmdefault}{\mddefault}{\updefault}{\color[rgb]{0,0,0}$\un$}%
}}}}
\end{picture}%
} \star_0 \figeps{nil} \right)
\star_1 \left( \raisebox{-1.25mm}{\begin{picture}(0,0)%
\includegraphics{2.pstex}%
\end{picture}%
\setlength{\unitlength}{4144sp}%
\begingroup\makeatletter\ifx\SetFigFont\undefined%
\gdef\SetFigFont#1#2#3#4#5{%
  \reset@font\fontsize{#1}{#2pt}%
  \fontfamily{#3}\fontseries{#4}\fontshape{#5}%
  \selectfont}%
\fi\endgroup%
\begin{picture}(106,204)(1118,1097)
\put(1171,1177){\makebox(0,0)[b]{\smash{{\SetFigFont{6}{7.2}{\rmdefault}{\mddefault}{\updefault}{\color[rgb]{0,0,0}$\deux$}%
}}}}
\end{picture}%
} \star_0 \figeps{cons} \right)
\star_1 \figeps{cons} 
\: = \:
\raisebox{-3mm}{\raisebox{-1.25mm}{\input{liste-2-1.pstex_t}}}.
$$

\noindent The value $\sem{\figeps{sort}} \left(\ul{[2;1]} \right)$ is, by definition, the unique normal form of the $2$-path $\ul{[2;1]}\star_1\figeps{sort}$. Figure~\ref{Figure:Computation} presents a normalizing $3$-path, obtained by $\star_2$-composition of smaller $3$-paths, where we have given self-explanatory "names" to the involved $3$-cells, without further explanations. After computation, one gets the expected $\sem{\figeps{sort}} \left( \ul{[2;1]} \right) = \ul{[1;2]}$ as the target of this $3$-path.
\end{exa}

\subsection{Polygraphic programs are Turing-complete}
\label{Subsection:TuringComplete}

This completeness result is not a surprising one. Indeed, one could argue, for instance, that polygraphic programs simulate term rewriting systems, a Turing-complete model of computation. Our proof, similar to the one concerning interaction nets~\cite{Lafont90}, prepares for the encoding of Turing machines with clocks, used for Theorem~\ref{Theorem:PTIME}.

\begin{defi}
A \emph{Turing machine} is a family $\Mr=\left( \Sigma, Q, q_0, q_f, \delta \right)$ made of:
\begin{enumerate}[$\bullet$]
\item A finite set $\Sigma$, called the \emph{alphabet}; one denotes by $\overline{\Sigma}$ its extension with a new element, denoted by~$\sharp$ and called the \emph{blank character}.
\item A finite set $Q$, whose elements are called \emph{states}, two distinguished elements~$q_0$, the \emph{initial state}, and $q_f$, the \emph{final state}.
\item A map $\delta: (Q - \ens{q_f} ) \times \overline{\Sigma} \fl Q \times \overline{\Sigma} \times\ens{L,R}$, called the \emph{transition function}, where $\ens{L,R}$ is any set with two elements.
\end{enumerate}

\noindent A \emph{configuration} of $\Mr$ is an element $(q,a,w_l,w_r)$ of the product set $Q\times\overline{\Sigma}\times\mon{\overline{\Sigma}}\times\mon{\overline{\Sigma}}$: here~$q$ is the current state of the machine, $a$ is the currently read symbol, $w_l$ is the word at the left-hand side of $a$ and~$w_r$ is the word at the right-hand side of $a$. For further convenience, the word~$w_l$ is written in reverse order, so that its first letter is the one that is immediately at the left of $a$. 

The \emph{transition relation} of $\Mr$ is the binary relation denoted by~$\fl_{\Mr}$ and defined on the set of configurations of~$\Mr$ as follows, where~$e$ denotes the neutral element of~$\mon{\Sigma}$:
\begin{enumerate}[$\bullet$]
\item If $\delta(q_1,a)=(q_2,c,L)\:$ then
$\:\left\{
\begin{array}{l c l}
\left( q_1, a, e, w_r \right) &\fl_{\Mr}& \left( q_2, \sharp, e, c w_r \right), \\
\left( q_1, a, b w_l, w_r \right) &\fl_{\Mr}& \left( q_2, b, w_l, c w_r \right).
\end{array}
\right.$
\item If $\delta(q_1,a)=(q_2,c,R)\:$ then 
$\:\left\{
\begin{array}{l c l}
\left( q_1, a, w_l, e \right) &\fl_{\Mr}& \left( q_2, \sharp, c w_l, e \right), \\
\left( q_1, a, w_l, bw_r \right) &\fl_{\Mr}& \left( q_2, b, c w_l, w_r \right).
\end{array}
\right.$
\end{enumerate} 

\noindent One denotes by $\fl_{\Mr}^*$ the reflexive and transitive closure of~$\fl_{\Mr}$. Let $f:\mon{\Sigma}\fl\mon{\Sigma}$ be a map. One says that $\Mr$ \emph{computes $\mathit{f}$} when, for any $w$ in $\mon{\Sigma}$, there exists a configuration of the shape $(q_f,a,v,f(w))$ such that $(q_0,\sharp,e,w)\fl_{\Mr}^*(q_f,a,v,f(w))$ holds (in that case, this final configuration is unique). 
\end{defi}

\begin{thm}\label{Theorem:TuringComplete}
Polygraphic programs form a Turing-complete model of computation.
\end{thm}

\proof
We fix a Turing machine $\Mr=\left(\Sigma,Q,q_0,q_f,\delta\right)$ and a map $f$ computed by $\Mr$. From this Turing machine, we build the following polygraphic program $\Pr(\Mr)$:

\begin{enumerate}[(1)]

\item It has one $1$-cell $\mathtt{w}$, standing for the type of words over $\Sigma$.

\item Apart from the three structure $2$-cells, its $2$-cells consist of:

\begin{enumerate}[(a)]
\item Constructor $2$-cells: $\figeps{nil}:\ast\dfl\mathtt{w}$, for the empty word, plus one $\raisebox{-1.25mm}{\begin{picture}(0,0)%
\includegraphics{a.pstex}%
\end{picture}%
\setlength{\unitlength}{4144sp}%
\begingroup\makeatletter\ifx\SetFigFont\undefined%
\gdef\SetFigFont#1#2#3#4#5{%
  \reset@font\fontsize{#1}{#2pt}%
  \fontfamily{#3}\fontseries{#4}\fontshape{#5}%
  \selectfont}%
\fi\endgroup%
\begin{picture}(123,204)(344,557)
\put(406,637){\makebox(0,0)[b]{\smash{{\SetFigFont{6}{7.2}{\rmdefault}{\mddefault}{\updefault}{\color[rgb]{0,0,0}$a$}%
}}}}
\end{picture}%
}:\mathtt{w}\dfl \mathtt{w}$ for each $a$ in $\Sigma$.
\item Function $2$-cells: $\figeps{sort}:\mathtt{w}\dfl\mathtt{w}$, for the map $f$, plus one $\raisebox{-0.5mm}{\raisebox{-1.25mm}{\input{step-q-a.pstex_t}}}:\mathtt{w}\star_0\mathtt{w}\dfl\mathtt{w}$ for each pair $(q,a)$ in $Q\times\bar{\Sigma}$, for the behaviour of the Turing machine.
\end{enumerate}

\item Its $3$-cells are the structure ones, plus the following computation $3$-cells -- the first one initializes the computation, the four subsequent families simulate the transitions of the Turing machine and the final cell starts the computation of the result:

\medskip
$\qquad\raisebox{-1.25mm}{\input{3-cellules-turing.pstex_t}}$
\end{enumerate}

\medskip
\noindent One checks that $\sem{\mathtt{w}}\simeq\mon{\Sigma}$ through $\ul{e}=\figeps{nil}$ and $\ul{aw}=\ul{w}\star_1\raisebox{-1.25mm}{\begin{picture}(0,0)%
\includegraphics{a.pstex}%
\end{picture}%
\setlength{\unitlength}{4144sp}%
\begingroup\makeatletter\ifx\SetFigFont\undefined%
\gdef\SetFigFont#1#2#3#4#5{%
  \reset@font\fontsize{#1}{#2pt}%
  \fontfamily{#3}\fontseries{#4}\fontshape{#5}%
  \selectfont}%
\fi\endgroup%
\begin{picture}(123,204)(344,557)
\put(406,637){\makebox(0,0)[b]{\smash{{\SetFigFont{6}{7.2}{\rmdefault}{\mddefault}{\updefault}{\color[rgb]{0,0,0}$a$}%
}}}}
\end{picture}%
}$. Then, to every configuration $(q,a,w_l,w_r)$, one associates the $2$-path $\ul{(q,a,w_l,w_r)} \:=\: \left( \ul{w_l} \star_0 \ul{w_r} \right) \star_1 \raisebox{-0.5mm}{\raisebox{-1.25mm}{\input{step-q-a.pstex_t}}}$. The four cases in the definition of the transition relation of $\Mr$ are in one-to-one correspondence with the four middle families of $3$-cells of the polygraph $\Pr(\Mr)$. Hence the following equivalence holds:
$$
(q,a,w_l,w_r) \:\fl_{\Mr}^*\: (q',a',w'_l,w'_r) \quad\text{if and only if}\quad \ul{(q,a,w_l,w_r)} \:\tfl\: \ul{(q',a',w'_l,w'_r)}.
$$

\noindent Finally, let us fix a $w$ in $\mon{\Sigma}$. Since $\Mr$ computes $f$, there exists a unique configuration $(q_f,a,v,f(w))$, such that $(q_0,\sharp,e,w)\fl_{\Mr}^*(q_f,a,v,f(w))$ holds. As a consequence, $\ul{w}\star_1\figeps{sort}$ has a unique normal form, so that the following equalities hold, yielding $\sem{\figeps{sort}}=f$:
$$
\sem{\figeps{sort}} \left(\ul{w}\right) 
\:=\: 
\sem{\raisebox{-0.5mm}{\raisebox{-1.25mm}{\input{step-q0-a.pstex_t}}}} \left(\figeps{nil} \star_0 \ul{w} \right)
\:=\:
\sem{\raisebox{-0.5mm}{\raisebox{-1.25mm}{\input{step-qf-a.pstex_t}}}} \left( \ul{v} \star_0 \ul{f(w)} \right)
\:=\: 
\ul{f(w)}.\eqno{\qEd} 
$$

\section{Polygraphic interpretations} 
\label{Section:Interpretations}

\noindent Here, we present general results about information that can be recovered from functorial and differential interpretations of $3$-polygraphs. 

\subsection{Functorial interpretations}
\label{Subsection:Currents}

\begin{defi}\label{Definition:Currents}
A \emph{functorial interpretation} of a $3$-polygraph $\Pr$ is a pair $\phi=(\phi_1,\phi_2)$ consisting of:
\begin{enumerate}[(1)]
\item a map $\phi_1$ sending every $1$-path $u$ of size $n$ to a non-empty part of $(\Nb-\ens{0})^n$;
\item a map $\phi_2$ sending every $2$-path $f:u\dfl v$ to a monotone map from $\phi_1(u)$ to $\phi_1(v)$.
\end{enumerate}

\noindent The following equalities, called \emph{functorial relations}, must be satisfied:
\begin{enumerate}[$\bullet$]
\item if $u$ is a degenerate $2$-path, then $\phi_2(u)$ is the identity of $\phi_1(u)$;
\item if $u$ and $v$ are $0$-composable $1$-paths, then $\phi_1(u\star_0 v) = \phi_1(u)\times\phi_1(v)$ holds;
\item if $f$ and $g$ are $0$-composable $2$-paths, then $\phi_2(f\star_0 g) = \phi_2(f)\times\phi_2(g)$ holds;
\item if $f$ and $g$ are $1$-composable $2$-paths, then $\phi_2(f\star_1 g) = \phi_2(g)\circ\phi_2(f)$ holds.
\end{enumerate}
\end{defi}
  
\noindent One simply writes $\phi$ for both $\phi_1$ and $\phi_2$. Intuitively, for every $2$-cell $\figeps{fonction}$, the map $\phi(\figeps{fonction})$ tells us how~$\figeps{fonction}$, seen as a circuit gate, transmits currents downwards. In practice, one computes the value of a current interpretation on a $2$-path by computing it on the $2$-cells it contains and assembling them in an intuitive way. The following result formalizes this fact.

\begin{lem}\label{Lemma:UniversalCurrents}
A functorial interpretation of a $3$-polygraph $\Pr$ is entirely and uniquely defined by its values on the $1$-cells and $2$-cells of $\Pr$.
\end{lem}

\proof Using the functorial relations, one checks that a functorial
interpretation takes the same values on both sides of the relations of
associativity, local units and exchange on $2$-paths: this property
comes from the fact that set-theoretic maps satisfy these same
relations. Then the functorial relations give the values of a current
interpretation on $2$-paths of size $n+1$ from its values on $2$-paths
of size $k\leq n$.  \qed

\noindent A direct consequence of Lemma~\ref{Lemma:UniversalCurrents} is that, when one wants to introduce a functorial interpretation, one only has to give its values on the $1$-cells and on the $2$-cells.

\begin{exa}\label{Example:CartesianCurrents}
  Let $\Pr$ be a polygraphic program with no constructor $2$-cell and
  no function $2$-cell. Then, given a non-empty part $\phi(\xi)$ of
  $\Nb-\ens{0}$ for every $1$-cell $\xi$, the following values extend
  $\phi$ into a functorial interpretation of $\Pr$:
$$
\phi \left( \figeps{tau}_{\xi,\zeta} \right)(x,y) \:=\: (y,x) 
\qquad\text{and}\qquad 
\phi \left( \figeps{delta}_{\xi} \right) (x) \:=\: (x,x).
$$
  Let us note that every functorial interpretation $\phi$ must send
  the $0$-cell $\ast$ to some single-ele\-ment part
  of~$\Nb-\ens{0}$. Hence, it must assign each
  $\figeps{epsilon}_{\xi}$ to the only map from~$\phi(\xi)$
  to~$\phi(\ast)$.
\end{exa}

\begin{exa}\label{Example:DivisionCurrents}
  The following values extend the ones of
  Example~\ref{Example:CartesianCurrents} into a functorial
  in\-ter\-pretation of the polygraphic program~$\Dr$ of division:
$$
\phi(\mathtt{n}) = \Nb-\ens{0}, \qquad \phi(\figeps{nil}) = 1, \qquad \phi(\figeps{succ})(x) = x+1, 
$$
$$
\phi(\figeps{fonction-2-b})(x,y) = \phi(\figeps{fonction-2})(x,y) = x.
$$
\end{exa}

\begin{exa}\label{Example:FusionSortCurrents}
For the polygraphic program $\Fr$ of fusion sort, we extend the
functorial in\-ter\-pretation of Example~\ref{Example:CartesianCurrents}
with the following values, where $\roundup{\cdot}$ and
$\rounddown{\cdot}$ stand for the rounding functions, respectively by
excess and by default:
$$
\phi(\texttt{n}) = \ens{1}, \qquad \phi(\texttt{l}) = 2\Nb+1, \qquad \phi(\raisebox{-1.25mm}{\begin{picture}(0,0)%
\includegraphics{n.pstex}%
\end{picture}%
\setlength{\unitlength}{4144sp}%
\begingroup\makeatletter\ifx\SetFigFont\undefined%
\gdef\SetFigFont#1#2#3#4#5{%
  \reset@font\fontsize{#1}{#2pt}%
  \fontfamily{#3}\fontseries{#4}\fontshape{#5}%
  \selectfont}%
\fi\endgroup%
\begin{picture}(139,174)(887,1097)
\put(957,1188){\makebox(0,0)[b]{\smash{{\SetFigFont{6}{7.2}{\rmdefault}{\mddefault}{\updefault}{\color[rgb]{0,0,0}$n$}%
}}}}
\end{picture}%
}) = \phi(\figeps{nil}) = 1, \qquad \phi(\figeps{cons})(x,y) = x+y+1, 
$$
$$
\phi(\figeps{sort})(x) = x, \qquad \phi(\figeps{merge})(x,y) = x+y-1, \qquad \phi(\figeps{split})(2x+1) = \left( 2\cdot\roundup{\frac{x}{2}}+1,\: 2\cdot\rounddown{\frac{x}{2}}+1 \right).
$$
\end{exa}

\begin{exa}\label{Example:FunctorialNorm}
  Let $\Pr$ be a polygraphic program. One denotes by $\nu$ the
  functorial interpreta\-tion on the subpolygraph $\mon{\Pr_2^C}$
  defined, for every $1$-cell $\xi$, by $\nu(\xi)=\Nb-\ens{0}$ and,
  for every constructor $2$-cell~$\figeps{phi}$ with arity~$n$, by:
$$
\nu(\figeps{phi})(x_1,\dots,x_n) \:=\: x_1+\dots+x_n+1.
$$
  One checks that $\nu(t)=\norm{t}$ holds for every value $t$ with
  coarity $1$. Thus, given values~$t_1$, $\dots$, $t_n$ with coarity
  $1$, the following equality holds in $\Nb^n$:
$$
\nu(t_1\star_0\dots\star_0 t_n) \: = \: \big( \norm{t_1},\dots,\norm{t_n} \big).
$$ 
  We use the functorial interpretation $\nu$ to describe the size of
  arguments of a function.
\end{exa}

\begin{lem}\label{Lemma:ContextualCurrents}
  Let $\phi$ be a functorial interpretation of a $3$-polygraph
  $\Pr$. Let $f$, $g$, $h$ and $k$ be $2$-paths such that
  $\phi(f)\leq\phi(g)$ and $\phi(h)\leq\phi(k)$ hold. Then, for every
  $i\in\ens{0,1}$ such that $f\star_i h$ is defined, the inequality
  $\phi(f\star_i h) \leq \phi(g\star_i k)$ is satisfied.
\end{lem}

\proof
  One has:
$$
\phi(f\star_0 h) \:=\: \phi(f)\times\phi(h) \:\leq\: \phi(g)\times\phi(k) \:=\: \phi(g\star_0 k).
$$
  Indeed, the two equalities are given by the functorial relations
  that $\phi$ satisfies, while the middle inequality comes from the
  hypotheses and the fact that one uses a product order. Then one has:
$$
\phi(f\star_1 h) \:=\: \phi(h)\circ\phi(f) \:\leq\: \phi(h)\circ\phi(g) \:\leq\: \phi(k)\circ\phi(g) \:=\: \phi(g\star_1 k).
$$
  The equalities come from the functorial relations; the first
  inequality uses the hypothesis $\phi(f)\leq\phi(g)$ and the fact
  that $\phi(h)$ is monotone; the second inequality uses
  $\phi(h)\leq\phi(k)$ and the fact that maps are compared pointwise.
  \qed

\subsection{Compatible functorial interpretations}
\label{Subsection:CompatibleCurrents}

\begin{defi}\label{Definition:CompatibleCurrents}
Let $\phi$ be a functorial interpretation of a
$3$-polygraph~$\Pr$. For every $3$-cell~$\alpha$ of $\Pr$, one says
that $\phi$ is \emph{compatible with $\mathit{\alpha}$} when the
inequality $\phi(s_2\alpha)\geq\phi(t_2\alpha)$ holds. One says
that~$\phi$ is \emph{compatible} when it is compatible with every
$3$-cell of $\Pr$.
\end{defi}

\begin{exa}
The functorial interpretations given in
Examples~\ref{Example:DivisionCurrents}
and~\ref{Example:FusionSortCurrents} are compatible with all the
$3$-cells of the corresponding $3$-polygraph. We will see later that
the values they take on structure $2$-cells ensure that they are
compatible with all the structure $3$-cells. Concerning the
computation $3$-cells, let us consider, for example, the third one
associated to the sort function $2$-cell~$\figeps{sort}$. For the
source, one gets:
\begin{align*}
\phi \left( \raisebox{-3mm}{\figeps{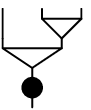}}  \right) (1,1,2x+1) \quad
&=\quad \phi \left( \raisebox{-1.5mm}{\figeps{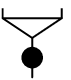}} \right) \big( 1, \phi(\figeps{cons})(1,2x+1) \big) \\
&=\quad \phi(\figeps{sort}) \circ \phi(\figeps{cons}) ( 1, 2x+3 ) \\
&=\quad \phi(\figeps{sort}) (2x+5) \\
&= \quad 2x+5.
\end{align*}

\noindent Now, for the target, going quicker:
\begin{displaymath}
\phi \left( \raisebox{-6mm}{\figeps{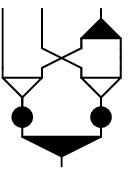}} \right) (1,1,2x+1) 
\: = \:
\phi(\figeps{merge}) \big( 2\cdot\roundup{x/2} + 3, \:2\cdot\rounddown{x/2} + 3 \big)
\: = \: 
2x+5.
\end{displaymath}
\end{exa}

\smallskip
\begin{prop}\label{Proposition:CompatibleCurrents}
Let $\phi$ be a compatible functorial interpretation of a polygraphic program. Then, for every $3$-path $F$, the inequality $\phi(s_2 F)\geq\phi(t_2 F)$ holds.
\end{prop}

\proof
We proceed by induction on the size of $3$-paths. If $F$ is a degenerate $3$-path, then $s_2 F=t_2 F$ holds and, thus, so does $\phi(s_2 F)=\phi(t_2 F)$. 

Let us assume that $F$ is an elementary $3$-path. Then one
decomposes~$s_2F$ and~$t_2F$, using a $3$-cell $\alpha$, $2$-paths
$f$, $g$ and $1$-paths $u$, $v$, yielding:
$$
\phi(s_2 F) \:=\: \phi \big( f\star_1 (u\star_0 s_2\alpha\star_0 v) \star_1 g \big) 
\qquad\text{and}\qquad
\phi(t_2 F) \:=\: \phi \big( f\star_1 (u\star_0 t_2\alpha\star_0 v) \star_1 g \big).
$$

\noindent The functorial interpretation $\phi$ is compatible with
$\alpha$, hence $\phi(s_2\alpha)\geq \phi(t_2\alpha)$ holds. Then one
applies Lemma~\ref{Lemma:ContextualCurrents} four times to get
$\phi(s_2 F)\geq\phi(t_2 F)$.

Now, let us fix a non-zero natural number $N$ and assume that the
property holds for every $3$-path of size $N$. Let us consider a
q$3$-path $F$ of size $N+1$. Then one decomposes $F$ into $G\star_2 H$
where $G$ is a $3$-path of size $N$ and $H$ is an elementary
$3$-path. One concludes using the induction hypothesis on~$G$ and the
previous case on~$H$.  \qed

\subsection{Differential interpretations}
\label{Subsection:Heat}

In this work, we use differential interpretations as an abstraction of
"heats", but also, later, to define the property of conservativeness
on "currents".  For this reason, we introduce the following
abstraction:

\begin{defi}
A \emph{(strictly) ordered commutative monoid} is an ordered set
$(M,\preceq)$ equip\-ped with a commutative monoid structure $(+,0)$
such that $+$ is (strictly) monotone in both arguments.
\end{defi}

\begin{exa}
Concretely, in what follows, we consider $\Nb$ equipped with its
natural order and either the addition (strict case) or the maximum map
(non-strict case), both with $0$ as neutral element.
\end{exa}

\begin{defi}\label{Definition:Heat}
Let $M$ be an ordered commutative monoid, let~$\Pr$ be a $3$-polygraph
and let~$\phi$ be a functorial interpretation of~$\Pr$. A
\emph{differential interpretation of~$\mathit{\Pr}$
  over~$\mathit{\phi}$ into $M$} is a map $\dr$ that sends each
$2$-path~$\figeps{fonction}$ of $\Pr$ with $1$-source $u$ to a
monotone map $\dr\figeps{fonction}$ from $\phi(u)$ to $M$, such that
the following conditions, called \emph{differential relations}, are
satisfied:
\begin{enumerate}[$\bullet$]
\item If $u$ is degenerate then $\dr u = 0$.
\item If $f$ and $g$ are $0$-composable then $\dr(f\star_0 g)(x,y) = \dr f (x) + \dr g(y)$ holds.
\item If $f$ and $g$ are $1$-composable then $\dr(f\star_1 g) = \dr f + \dr g\circ\phi(f)$ holds.
\end{enumerate}
\end{defi}

\noindent Intuitively, given a $2$-cell $\figeps{fonction}$, the map
$\dr\figeps{fonction}$ tells us how much heat it produces, when seen
as a circuit gate, depending on the intensities of incoming
currents. In order to compute the heat produced by a $2$-path, one
determines the currents that its $2$-cells propagate and, from those
values, the heat each one produces; then one sums up all these heats.

\begin{lem}\label{Lemma:UniversalPropertyHeat}
A differential interpretation of a polygraph $\Pr$ is entirely and
uniquely determined by its values on the $2$-cells of $\Pr$.
\end{lem}

\proof First, we prove that the differential relations imply that a
differential interpretation takes the same values on each side of the
relations of associativity, local units and exchange. For example, let
us check this for the exchange relation. For that, let us fix
$2$-paths $f$, $g$, $h$ and $k$ such that both $t_1(f)=s_1(h)$ and
$t_1(g)=s_1(k)$ are satisfied. We consider $x$ in $\phi(s_1(f))$ and
$y$ in $\phi(s_1(g))$ and, using the functorial relations of $\phi$
and the differential relations of $\dr$, we compute each one of the
following equalities in~$M$:
$$
\begin{array}{r c l}
\dr\big( (f\star_0 g) \star_1 (h\star_0 k) \big) (x,y)
&\:=\:& \big( \dr f(x) + \dr g(y) \big) + \big( \dr h\circ \phi(f)(x) + \dr k\circ \phi(g)(y) \big), \vspace{2mm} \\ 
\dr\big( (f\star_1 h) \star_0 (g\star_1 k) \big) (x,y)
&\:=\:& \big( \dr f(x) + \dr h \circ \phi(f)(x) \big) + \big( \dr g(y) + \dr k\circ\phi(g)(y) \big).
\end{array}
$$

\noindent One concludes using the associativity and commutativity
of~$+$ in~$M$. After that, one checks that the differential relations
determine the values of a differential interpretation on $2$-paths of
size $n+1$ from its values on $2$-paths of size $k\leq n$.  \qed

\noindent Lemma~\ref{Lemma:UniversalPropertyHeat} allows one to define
a differential interpretation by giving its values on $2$-cells.

\begin{exa}\label{Example:SizeHeat}
The \emph{trivial} functorial interpretation of a $3$-polygraph $\Pr$
sends every $1$-cell to some fixed one-element part $\ast$ of
$\Nb-\ens{0}$ and every $2$-path from $u$ to $v$ to the only possible
map from $\phi(u)\simeq\ast$ to $\phi(v)\simeq\ast$. Now, let us fix a
family $X$ of $2$-cells in $\Pr$. One can check that the map
$\norm{\cdot}_X$ is the differential interpretation of $\Pr$ over the
trivial interpretation and into $(\Nb,+,0)$, sending a $2$-cell
$\figeps{fonction}$ to $1$ if it is in $X$ and $0$ otherwise.
\end{exa}

\begin{exa}\label{Example:DivisionHeat}
We consider the differential interpretation of the division
polygraphic program $\Dr$, over the functorial interpretation given in
Example~\ref{Example:DivisionCurrents}, into $(\Nb,+,0)$, sending
every constructor and structure $2$-cell to zero and:
$$
\dr\figeps{fonction-2-b}(x,y) \:=\: y+1 
\qquad\text{and}\qquad
\dr\figeps{fonction-2}(x,y) \:=\: xy+x,
$$
\end{exa}

\begin{exa}\label{Example:FusionSortHeat}
For the polygraphic program $\Fr$ of fusion sort, we consider the
differential interpretation, over the functorial interpretation of
Example~\ref{Example:FusionSortCurrents}, into $(\Nb,+,0)$, sending
every constructor and structure $2$-cells to zero and:
$$
\dr\figeps{sort} (2x+1) = 2x^2 +1,\quad 
\dr\figeps{split} (2x+1) = \rounddown{x/2}+1, \quad
\dr\figeps{merge} (2x+1,2y+1) = 
\begin{cases}
1 &\text{if } xy=0,  \\
x+y &\text{otherwise}.
\end{cases}
$$
\end{exa}

\begin{lem}\label{Lemma:ContextualHeats}
Let $\Pr$ be a $3$-polygraph, with a differential interpretation
$\dr$, over a functorial interpretation $\phi$, into an ordered
commutative monoid $(M,+,0,\preceq)$. Let $f$, $g$, $h$, $k$ be
$2$-paths such that the inequalities $\phi(f)\leq\phi(g)$, $\dr
f\preceq \dr g$ and $\dr h\preceq \dr k$ hold. Then, for every
$i\in\ens{0,1}$ such that $f\star_i h$ is defined, one has
$\dr(f\star_i h) \preceq \dr(g\star_i k)$. Moreover, when $M$ is
strictly ordered and either $\dr f\prec \dr g$ or $\dr h\prec \dr k$
hold, one has $\dr(f\star_i h) \prec \dr(g\star_i k)$.
\end{lem}

\proof
One computes, for $x\in\phi(s_1 f)$ and $y\in\phi(s_1 h)$:
$$
\dr(f\star_0 h)(x,y) \:=\: \dr f (x) + \dr h (y) \:\preceq\: \dr g (x) + \dr k (y) \:=\: \dr (g\star_0 k)(x,y).
$$

\noindent Indeed, the two equalities are given by the differential
relations that $\dr$ satisfies; the inequality uses the hypotheses,
the fact that maps are compared pointwise and the monotony of
$+$. Moreover, if $+$ is strictly monotone and if one of $\dr f\prec
\dr g$ or $\dr h\prec \dr k$ holds, then the middle inequality is
strict. Now, one checks:
$$
\dr(f\star_1 h) \:=\: \dr f + \dr h \circ\phi(f) \:\preceq\: \dr g + \dr k\circ\phi(g) \:=\: \dr(g\star_1 k).
$$
The equalities come from the differential relations; the
inequality comes from the hypotheses $\dr f\preceq \dr g$, $\dr
h\preceq\dr k$ and $\phi(f)\leq\phi(g)$, plus the monotony of $\dr h$
and $+$ and the fact that maps are compared pointwise. When $+$ is
strictly monotone and when either $\dr f\prec \dr g$ or $\dr h\prec\dr
k$ hold, the middle inequality is strict.  \qed

\subsection{Compatible differential interpretations}
\label{Subsection:CompatibleHeats}

\begin{defi}
Let $\Pr$ be a $3$-polygraph equipped with a functorial interpretation
$\phi$ and a differential interpretation $\dr$ of $\Pr$ over $\phi$
and into an ordered commutative monoid $M$. For every $3$-cell
$\alpha$, one says that $\dr$ is \emph{compatible with
  $\mathit{\alpha}$} when $\dr(s_2\alpha)\succeq \dr(t_2\alpha)$
holds. It is said to be \emph{strictly compatible with
  $\mathit{\alpha}$} when $\dr (s_2\alpha)\succ \dr (t_2\alpha)$
holds. One says that $\dr$ is \emph{(strictly) compatible} when it is
with every $3$-cell of $\Pr$.
\end{defi}

\begin{exa}
The differential interpretations given in
Examples~\ref{Example:DivisionHeat} and~\ref{Example:FusionSortHeat}
are com\-patible with every structure $3$-cell and strictly compatible
with every computation $3$-cell of their $3$-polygraph.

Indeed, in the source and the target of every structure $3$-cell
$\alpha$, only constructor and structure $2$-cells appear. The
considered differential interpretations sends these to zero, yielding
$\dr (s_2\alpha)=\dr(t_2\alpha)=0$.

For an example of compatibility with a computation $3$-cell, we consider the third $3$-cell of the fusion sort function $2$-cell~$\figeps{sort}$. On one hand, one gets:
\begin{displaymath}
\dr\left( \raisebox{-3mm}{\figeps{2-source-sort-1}} \right) (1,1,2x+1) 
\: = \: \dr\figeps{sort} (2x+5) 
\: = \: 2(x+2)^2+1 
\: = \: 2x^2 + 8x + 9.
\end{displaymath}

\noindent And, on the other hand, one computes:
\begin{align*}
\dr\left( \raisebox{-6mm}{\figeps{2-but-sort}} \right) (1,1,2x+1) \quad
=&\:\; \begin{cases} 
\dr\figeps{sort} \big( 2\roundup{x/2} +3 \big) 
\: + \: \dr\figeps{sort} \big(  2\rounddown{x/2} +3 \big) \vspace{2mm} \\ 
\: + \: \dr\figeps{split} (2x+1) 
\: + \: \dr\figeps{merge} \big ( 2\roundup{x/2} +3, 2\rounddown{x/2} +3 \big) 
\end{cases} \\
=&\quad  2\cdot\big( \roundup{x/2} +1 \big)^2 
+ 2\cdot\big( \rounddown{x/2} +1 \big)^2 
+ x + \rounddown{x/2} + 4 \\
=&\quad 2\roundup{x/2}^2+2\rounddown{x/2}^2+x+4\roundup{x/2}+5\rounddown{x/2}+8 \\
\leq& \quad  2x^2 + 6x + 8.
\end{align*}
\end{exa}

\begin{prop}\label{Proposition:CompatibleHeats}
Let $\dr$ be a compatible differential interpretation of a polygraphic program~$\Pr$, over a compatible functorial interpretation $\phi$ and into an ordered commutative monoid~$M$. Then, for every $3$-path $F$, the inequality $\dr(s_2 F)\succeq\dr(t_2 F)$ holds. When $M$ is strictly ordered, $\dr$ is strictly compatible and~$F$ is non-degenerate, then $\dr(s_2 F)\succ\dr(t_2 F)$ also holds. Moreover, if $M$ is $\Nb$ equipped with addition, then $\tnorm{F}\leq\dr(s_2 F)-\dr(t_2 F)$ holds.
\end{prop}

\proof
We proceed by induction on the size of $3$-paths. If $F$ is a degenerate $3$-path, then one has $s_2 F=t_2 F$ and, thus, $\dr(s_2 F)=\dr(t_2 F)$ also. 

Let us assume that $F$ is an elementary $3$-path. We decompose $F$ using a $3$-cell~$\alpha$, $2$-paths~$f$,~$g$ and $1$-paths $u$, $v$, yielding:
$$
\dr(s_2 F) \:=\: \dr \big( f\star_1 (u\star_0 s_2\alpha\star_0 v) \star_1 g \big)
\qquad\text{and}\qquad
\dr(t_2 F) \:=\: \dr \big( f\star_1 (u\star_0 t_2\alpha\star_0 v) \star_1 g \big).
$$

\noindent By assumption, $\phi$ and $\dr$ are compatible with $\alpha$, hence $\phi(s_2\alpha)\geq \phi(t_2\alpha)$ and $\dr(s_2\alpha)\succeq\dr(t_2\alpha)$ hold. Then one applies Lemmas~\ref{Lemma:ContextualCurrents} and~\ref{Lemma:ContextualHeats} to get $\dr(s_2 F)\succeq\dr(t_2 F)$ and, when~$\dr$ is strictly compatible with the $3$-cell~$\alpha$, $\dr(s_2 F)\succ\dr(t_2 F)$. If $M$ is $\Nb$, this means:
$$
\dr(s_2 F)-\dr(t_2 F)\geq 1=\tnorm{F}.
$$
  
\noindent Finally, let us fix a non-zero natural number $N$ and assume that the property holds for every $3$-path of size $N$. Let us consider a $3$-path $F$ of size $N+1$. Then one decomposes $F$ into $G\star_2 H$ where $G$ is a $3$-path of size $N$ and $H$ is an elementary $3$-path. Then we apply the induction hypothesis to $G$ and the previous case to $H$ to conclude.  
\qed

\subsection{Conservative functorial interpretations} \label{Subsection:Conservativity}
Intuitively, the following definition gives a bound on all the intensities of currents that one can find in the vicinity of any $2$-cell inside a $2$-path.

\begin{defi}
Let $\Pr$ be a $3$-polygraph equipped with a functorial interpretation
$\phi$. One denotes by~$\dr_{\phi}$ the differential interpretation of
$\Pr$, over $\phi$ and into $(\Nb,\max,0)$, sending every
$2$-cell~$\figeps{fonction}$ with valence~$(m,n)$, \ie with arity~$m$
and coarity~$n$, to the following map from
~$\phi(s_1\figeps{fonction})$ to $\Nb$:
$$
\dr_{\phi} \figeps{fonction} \:=\: \max\ens{ \mu_m,\: \mu_n\circ\phi(\figeps{fonction})},
$$
  
\noindent \ie $\dr_{\phi}\figeps{fonction}(x_1,\dots,x_m) = \max\ens{x_1,\dots,x_m,y_1,\dots,y_n}$, if $(y_1,\dots,y_n) = \phi(\figeps{fonction})(x_1,\dots,x_m)$. For every $3$-cell $\alpha$ of $\Pr$, one says that $\phi$ is \emph{conservative on $\mathit{\alpha}$} when $\dr_{\phi}$ is compatible with $\alpha$. One says that~$\phi$ is \emph{conservative} when it is conservative on every $3$-cell of $\Pr$, \ie when $\dr_{\phi}$ is compatible.
\end{defi}

\begin{exa}
  The functorial interpretations of
  Examples~\ref{Example:DivisionCurrents}
  and~\ref{Example:FusionSortCurrents} are conservative. Indeed, we
  shall see later that their values on structure and constructor
  $2$-cells ensure that they are conservative on structure
  $3$-paths. Let us check conservativeness on, for example, the last
  computation $3$-cell of the sort function $2$-cell~$\figeps{sort}$:
\begin{align*}
\dr_{\phi}\left( \raisebox{-3mm}{\figeps{2-source-sort-1}} \right) (1,1,2x+1)
\:&=\: \max\big\{ 1,\: 2x+1,\: 2x+2,\: 2x+3 \big\} \\
\:&=\: 2x+3 \\
\:&=\: \max\big\{ 1,\: 2x+1,\: 2\cdot\rounddown{x/2}+1,\: 2\cdot\roundup{x/2}+1,\\
  &\qquad 2\cdot\rounddown{x/2}+2,\: 2\cdot\roundup{x/2}+2,\: 2x+3 \big\} \\
\:&=\: \dr_{\phi}\left( \raisebox{-6mm}{\figeps{2-but-sort}} \right) (1,1,2x+1).
\end{align*}
\end{exa}

\noindent When a functorial interpretation is both compatible and conservative, the intensities of currents inside $2$-paths do not increase during computations.

\begin{prop}\label{Proposition:ConservativeCurrents}
Let $\phi$ be a compatible and conservative functorial interpretation of a polygraphic program. Then, for every $3$-path $F$, the inequality $\dr_{\phi}(s_2 F)\geq\dr_{\phi}(t_2 F)$ holds.
\end{prop}

\proof
By definition of conservativeness and using Proposition~\ref{Proposition:CompatibleHeats} on $\dr_{\phi}$.
\qed

\subsection{Polygraphic interpretations}
\label{Subsection:Interpretations}

\begin{defi}\label{Definition:PolygraphicInterpretations}
A \emph{polygraphic interpretation} of a $3$-polygraph $\Pr$ is a pair
$(\phi,\dr)$ made of a functorial interpretation $\phi$ of $\Pr$,
together with a differential interpretation $\dr$ of $\Pr$ over $\phi$
and into $(\Nb,+,0)$. In that case, $\phi$ and $\dr$ respectively are
the \emph{functorial part} and the \emph{differential part} of
$(\phi,\dr)$.

Let us fix a $3$-cell~$\alpha$. A polygraphic interpretation
$(\phi,\dr)$ is \emph{compatible (with $\mathit{\alpha}$)} when
both~$\phi$ and~$\dr$ are. It is \emph{strictly compatible (with
  $\mathit{\alpha}$)} when~$\phi$ is compatible with $\alpha$ and
$\dr$ is strictly compatible (with $\alpha$). It is \emph{conservative
  (on $\alpha$)} when $\phi$ is.
\end{defi}

\begin{exa}
The functorial and differential interpretations we have built on the
poly\-graphic programs of division and of fusion sort are two examples
of polygraphic interpreta\-tions that are conservative, compatible with
every structure $3$-cell and strictly compatible with every
computation $3$-cell.

Let us consider the trivial functorial interpretation and the
differential interpretation $\norm{\cdot}_X$ over it, for some family
$X$ of $2$-cells. They form a polygraphic interpretation that is
conservative but that has no general reason to be compatible with any
$3$-cell.
\end{exa}

\noindent We recall the following theorem:

\begin{thm}[\cite{Guiraud06jpaa}]\label{Theorem:Termination}
If a $3$-polygraph has a polygraphic interpretation which is strictly
compatible with all of its $3$-cells, then it terminates.
\end{thm}

\proof By application of
Proposition~\ref{Proposition:CompatibleHeats}, one knows that $\dr
(s_2 F)>\dr(t_2 F)$ holds for every elementary $3$-cell
$F$. Furthermore, these are maps with values into $\Nb$. Since there
is no infinite strictly decreasing sequence of such maps for the
pointwise order, one concludes that $\Pr$ must terminate.  \qed

\noindent In what follows, we use Theorem~\ref{Theorem:Termination} in several steps, thanks to the following result:

\begin{prop}\label{Proposition:RelativeTermination}
Let $\Pr$ be a $3$-polygraph and let $X$ be a set of $3$-cells of
$\Pr$. Let us assume that there exists a compatible polygraphic
interpretation on $\Pr$ whose restriction to $X$ is strictly
compatible. Then $\Pr$ terminates if and only if $\Pr - X$ does.
\end{prop}

\proof If $\Pr$ terminates, its reduction graph has no infinite
path. Since it contains the reduction graph of the $3$-polygraph
$\Pr-X$, the latter does not have any infinite path either. Hence
$\Pr-X$ terminates.

  Conversely, let us assume that $\Pr$ does not terminate. Then there
  exists an infinite sequence $(F_n)_{n\in\Nb}$ of elementary
  $3$-paths in $\Pr$ such that, for every $n\in\Nb$, $F_n$ and
  $F_{n+1}$ are composable. The polygraphic interpretation is
  compatible, hence one can apply
  Proposition \ref{Proposition:CompatibleHeats} to get the following
  infinite sequence of inequalities in $\Nb$:
$$
\dr(s_2 F_0) \:\geq\: \dr(t_2 F_0) \:=\: \dr(s_2 F_1) \:\geq\: (\cdots) \: = \: \dr (s_2 F_n) \:\geq\: \dr(t_2 F_n) \:=\: \dr(s_2 F_{n+1}) \:\geq\: (\cdots)
$$
  Furthermore, for every $n\in\Nb$ such that $F_n\in\mon{X}$, one has
  a strict inequality $\dr(s_2 F_n)>\dr(t_2 F_n)$, since the
  polygraphic interpretation is strictly compatible with every
  $3$-cell of $X$. Hence, there are only finitely many $n$ in $\Nb$
  such that $F_n$ is in $\mon{X}$: otherwise, one could extract, from
  $(\dr(s_2 F_n))_{n\in\Nb}$, an infinite, strictly decreasing
  sequence of maps with values in~$\Nb$. Thus, there exists some
  $n_0\in\Nb$ such that $(F_n)_{n\geq n_0}$ is an infinite path in the
  reduction graph of~$\Pr-X$: this means that $\Pr-X$ does not
  terminate.  
\qed

\begin{exa}
  Let us consider the polygraphic programs for division and fusion
  sort, given in Examples~\ref{Example:Division}
  and~\ref{Example:FusionSort}. We have seen that each one admits a
  compatible polygraphic interpretation that is strictly compatible
  with their computation $3$-cells. Furthermore, as proved later, the
  structure $3$-cells, alone, terminate. Thus
  Proposition~\ref{Proposition:RelativeTermination} gives the
  termination of both polygraphic programs.
\end{exa}

\noindent Actually, in what comes next, we produce a standard differential
  interpretation that is strictly compatible with structure
  $3$-cells. However, in general, it is not compatible, even in a
  non-strict way, with computation $3$-cells: informally, each
  application of such a cell can increase the "structure heat". The
  purpose of the rest of this section is to bound this potential
  augmentation.

\begin{lem}\label{Lemma:HeatBound}
  Let $\Pr$ be a $3$-polygraph equipped with a polygraphic
  interpretation $(\phi,\dr)$. Then, for every $2$-path $f$ in $\Pr$
  and every $x$ in $\phi(s_1 f)$, the following inequality holds in
  $\Nb$:
$$
\dr f(x) \:\leq\: \sum_{\raisebox{-.75mm}{\smallfigeps{fonction}}\in\Pr_2} \norm{f}_{\smallfigeps{fonction}} \cdot \dr\figeps{fonction} \: \big( \: \dr_{\phi}f(x), \: \dots, \: \dr_{\phi}f(x)\: \big).
$$
\end{lem}

\begin{rem}
  Let us note that we apply $\dr\figeps{fonction}$ to arguments
  $\dr_{\phi}f(x)$ that are not necessarily in its domain. In that
  case, one considers an extension of $\dr\figeps{fonction}$ sending
  $x$ to $\dr\figeps{fonction}(y)$, where~$y$ is the maximum element
  of the set $\phi(s_1\figeps{fonction})$ that is below $x$.
\end{rem}

\proof 
  We proceed by induction on the size of the $2$-path $f$. Let us
  assume that $f$ is degenerate. Then one has
  $\norm{f}_{\smallfigeps{fonction}}=0$ for every $2$-cell
  $\figeps{fonction}$ and, since $\dr$ is a differential
  interpretation, $\dr f=0$. Hence both sides of the sought inequality
  are equal to $0$.

  Now, let us consider an elementary $2$-path $f$. One decomposes $f$
  into $u\star_0\figeps{fonction}\star_0 v$, where~$\figeps{fonction}$
  is a $2$-cell and $u$ and $v$ are $1$-paths. Then
  $\norm{f}_{\smallfigeps{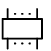}}$ is $1$ when
  $\figeps{fonction-b}$ is $\figeps{fonction}$ and $0$ otherwise. Let
  us fix $x$, $y$ and $z$ respectively in $\phi(u)$,
  $\phi(s_1\figeps{fonction})$ and $\phi(v)$. Using the differential
  relations of $\dr$ and $\dr_{\phi}$, one gets $\dr
  f(x,y,z)=\dr\figeps{fonction}(y)$ and
  $\dr_{\phi}f(x,y,z)=\dr_{\phi}\figeps{fonction}(y)$. If
  $\figeps{fonction}$ has valence $(m,n)$ and $y=(y_1,\dots,y_m)$, one
  uses the definition of $\dr_{\phi}\figeps{fonction}$ to get, for
  every $i\in\ens{1,\dots,m}$:
$$
\dr_{\phi}\figeps{fonction}(y) \:=\: \max\ens{\: \mu_m(y), \: \mu_n\circ\phi(\figeps{fonction})(y) \: } \:\geq\: y_i.
$$
  Then one computes:
\begin{align*}
\sum_{\smash{\raisebox{-.75mm}{\smallfigeps{fonction-b}}\in\Pr_2}}\, \norm{f}_{\smallfigeps{fonction-b}} \cdot \dr\figeps{fonction-b} \: \big( \: \dr_{\phi}f(x,y,z), \: \dots, \: \dr_{\phi}f(x,y,z) \: \big) 
\: &= \: \dr\figeps{fonction} \: \left(\: \dr_{\phi}\figeps{fonction}(y), \: \dots, \: \dr_{\phi}\figeps{fonction}(y) \: \right) \\
&\geq \: \dr\figeps{fonction} \: ( y_1,\dots,y_m) \\
&=\: \dr f(x,y,z).
\end{align*}
  Finally, let us fix a non-zero natural number $N$ and assume that
  the property holds for every $2$-path of size at most $N$. We
  consider a $2$-path $f$ of size $N+1$: there exists a decomposition
  $f=g\star_1 h$ where~$g$ and~$h$ are $2$-paths of size at most
  $N$. Then, using the differential relations of
  $\norm{\cdot}_{\smallfigeps{fonction}}$, for any
  $2$-cell~$\figeps{fonction}$, and of $\dr_{\phi}$, one gets:
$$
\norm{f}_{\smallfigeps{fonction}}\:=\:\norm{g}_{\smallfigeps{fonction}}+\norm{h}_{\smallfigeps{fonction}}
\qquad\text{and}\qquad
\dr_{\phi}(f) \:=\: \max\ens{\:\dr_{\phi}g,\:\dr_{\phi}h\circ\phi(g)\:}.
$$

\noindent We fix a $x$ in $\phi(s_1 f)$ and we compute:
\begin{align*}
\dr f (x) 
\quad &= \quad \dr (g\star_1 h) (x) \\
&= \quad \dr g (x) + \dr h\circ \phi(g)(x) \\
&\leq \quad \sum_{\raisebox{-.75mm}{\smallfigeps{fonction}}\in\Pr_2} \norm{g}_{\smallfigeps{fonction}} \cdot \dr\figeps{fonction} \: \big( \: \dr_{\phi}g(x), \:\dots, \:\dr_{\phi}g(x) \: \big) \\
&\qquad+ \sum_{\raisebox{-.75mm}{\smallfigeps{fonction}}\in\Pr_2} \norm{h}_{\smallfigeps{fonction}} \cdot \dr\figeps{fonction} \: \big( \: \dr_{\phi}h\circ\phi(g)(x), \:\dots, \:\dr_{\phi}h\circ\phi(g)(x) \: \big) \\
&\leq \quad \sum_{\raisebox{-.75mm}{\smallfigeps{fonction}}\in\Pr_2} \norm{g}_{\smallfigeps{fonction}} \cdot \dr\figeps{fonction} \: \big( \: \dr_{\phi}f(x), \:\dots, \:\dr_{\phi}f(x) \: \big) \\
&\qquad+ \sum_{\raisebox{-.75mm}{\smallfigeps{fonction}}\in\Pr_2} \norm{h}_{\smallfigeps{fonction}} \cdot \dr\figeps{fonction} \: \big( \: \dr_{\phi}f(x), \:\dots, \:\dr_{\phi}f(x) \: \big) 
\end{align*}

\noindent We factorize the right-hand side to conclude the proof:
\begin{align*}
\dr f (x) 
\quad &\leq \quad \sum_{\raisebox{-.75mm}{\smallfigeps{fonction}}\in\Pr_2} \left(\norm{g}_{\smallfigeps{fonction}}+\norm{h}_{\smallfigeps{fonction}}\right) \cdot \dr\figeps{fonction}\: \big( \: \dr_{\phi}f(x), \: \dots, \: \dr_{\phi}f(x) \: \big) \\
&=\quad \sum_{\raisebox{-.75mm}{\smallfigeps{fonction}}\in\Pr_2}
\norm{f}_{\smallfigeps{fonction}} \cdot \dr\figeps{fonction} \: \big(
\: \dr_{\phi}f(x), \: \dots, \: \dr_{\phi}f(x) \: \big).\rlap{\hbox to
  112 pt{\hfil\qEd}}
\end{align*}

\begin{prop}\label{Proposition:HeatIncrease}
Let $\Pr$ be a $3$-polygraph, let $\alpha$ be a $3$-cell of $\Pr$ and let $F$ be an elementary $3$-path in~$\mon{\alpha}$. One assumes that $\Pr$ is equipped with a polygraphic interpretation $(\phi,\dr)$ such that $\phi$ is compatible with and conservative on $\alpha$. Then, for every $x\in\phi(s_1 F)$, the following inequality holds in $\Zb$:
$$
\dr(t_2 F) (x) - \dr(s_2 F) (x) \:\leq\: \sum_{\raisebox{-.75mm}{\smallfigeps{fonction}}\in\Pr_2} \norm{t_2(\alpha)}_{\smallfigeps{fonction}} \cdot \dr\figeps{fonction} \: \big( \: \dr_{\phi}(s_2 F)(x), \: \dots, \: \dr_{\phi}(s_2 F)(x) \: \big).
$$
\end{prop}

\proof
Since $F$ is a $3$-path of size $1$ in $\mon{\alpha}$, one can decompose $s_2 F$ and $t_2 F$ as follows:
$$
\raisebox{-8mm}{\raisebox{-1.25mm}{\input{decomposition-reduction-f.pstex_t}}} \qquad\text{and}\qquad \raisebox{-8mm}{\raisebox{-1.25mm}{\input{decomposition-reduction-g.pstex_t}}} \: .
$$

\noindent Let us denote by $p$, $q$ and $m$ the respective sizes of $u$, $v$ and $s_1 F$. The map $\phi(f)$ takes its values in a part of $\Nb^{p+m+q}$: we decompose it into three maps denoted by $\phi_1(f)$, $\phi_2(f)$ and $\phi_3(f)$, with the same domain and respectively taking their values in parts of $\Nb^p$, $\Nb^m$ and~$\Nb^q$. Let us fix a $x\in\phi(s_1 F)$. The functorial and differential relations give:
$$
\dr (s_2 F) (x)
\:=\: \dr f (x) \:+\: \dr (s_2 \alpha) \circ \phi_2(f) (x) 
\:+\: \dr g \big( \: \phi_1(f)(x), \: \phi(s_2\alpha)\circ\phi_2(f)(x), \: \phi_3(f)(x) \: \big).
$$

\noindent With the same arguments, one gets the same decomposition for $\dr(t_2 F)$, with $s_2\alpha$ replaced by~$t_2\alpha$. Thus, the following holds in $\Zb$:
\begin{align*}
\dr(t_2 F)(x)-\dr(s_2 F)(x) 
&\:=\: \dr(t_2\alpha) \circ \phi_2(f)(x) - \dr(s_2\alpha) \circ \phi_2(f)(x) \\
&\quad\: +\: \dr g \: \big( \: \phi_1(f)(x), \: \phi(t_2\alpha)\circ\phi_2(f)(x), \: \phi_3(f)(x) \:\big) \\
&\quad\: -\: \dr g \: \big( \: \phi_1(f)(x), \: \phi(s_2\alpha)\circ\phi_2(f)(x), \: \phi_3(f)(x) \:\big).
\end{align*}

\noindent Let us prove that $\dr(t_2 F)(x) - \dr(s_2 F)(x) \leq \dr(t_2\alpha)\circ\phi_2(f)(x)$ holds. First, one has $\dr(s_2\alpha)\geq 0$. Moreover, $\phi$ is compatible with $\alpha$, which means that $\phi(s_2\alpha)\geq\phi(t_2\alpha)$ holds; since the map~$\dr g$ is monotone, the following holds in $\Nb$:
$$
\dr g \: \big(\: \phi_1(f)(x), \: \phi(s_2\alpha)\circ\phi_2(f)(x), \: \phi_3(f)(x) \: \big) 
\:\geq\:
\dr g \: \big( \: \phi_1(f)(x), \: \phi(t_2\alpha)\circ\phi_2(f)(x), \: \phi_3(f)(x) \: \big).
$$
  It remains to bound $\dr (t_2\alpha) \circ\phi_2(f)(x)$. One applies
  Lemma~\ref{Lemma:HeatBound} to $t_2(\alpha)$ to get:
$$
\dr(t_2\alpha)\circ\phi_2(f)(x) \: \leq \: 
\sum_{\raisebox{-.75mm}{\smallfigeps{fonction}}\in\Pr_2} \norm{t_2(\alpha)}_{\smallfigeps{fonction}} \cdot \dr\figeps{fonction}\: \big( \: \dr_{\phi}(t_2\alpha)\circ\phi_2(f)(x), \: \dots, \: \dr_{\phi}(t_2\alpha)\circ\phi_2(f)(x) \: \big).
$$
  By assumption, $\phi$ is conservative on $\alpha$, thus
  $\dr_{\phi}t_2(\alpha)\circ\phi_2(f)(x) \leq
  \dr_{\phi}s_2(\alpha)\circ\phi_2(f)(x)$ holds. Moreover, using the
  differential properties satisfied by $\dr_{\phi}$, one gets
  $\dr_{\phi}s_2(\alpha)\circ\phi_2(f)(x)\leq \dr_{\phi}(s_2 F)$. One
  concludes by invoking the monotony of~$\dr\figeps{fonction}$.  \qed

\section{Complexity of polygraphic programs}
\label{Section:Complexity}

\noindent In this section, we specialize polygraphic interpretations
to polygraphic programs to get information on their complexity. In
particular, we introduce additive polygraphic interpreta\-tions and use
them as an estimation of the size of values. This way, we give bounds
on the size of computations, with respect to the size of the
arguments. We conclude this work with a characterisation of a class of
polygraphic programs that compute exactly the \Ptime functions.

\subsection{Additive functorial interpretations and the size of values}
\label{Subsection:SpaceComplexity}

\begin{defi}\label{Definition:AdditiveCurrents}
Let $\Pr$ be a polygraphic program. One says that a functorial interpretation~$\phi$ of $\Pr$ is \emph{additive} when, for every constructor $2$-cell $\figeps{phi}$ of arity $n$, there exists a non-zero natural number $c_{\smallfigeps{phi}}$ such that, for every $(x_1,\dots,x_n)$ in $\phi(s_1\figeps{phi})$, the following equality holds in $\Nb$:
$$
\phi(\figeps{phi})(x_1,\dots,x_n) \:=\: x_1+\dots+x_n+c_{\smallfigeps{phi}}.
$$ 

\noindent In that case, one denotes by $\gamma$ the greatest of these numbers, \ie:
$$
\gamma \:=\: \max \ens{\: c_{\smallfigeps{phi}}, \: \figeps{phi}\in\Pr_2^C }.
$$ 

\noindent A polygraphic interpretation is \emph{additive} when its functorial part is.
\end{defi}

\begin{exa}
The functorial interpretations we have built for the polygraphic programs~$\Dr$ and~$\Fr$ are additive. In both cases, $\gamma$ is $1$.
\end{exa}

\begin{lem}\label{Lemma:AdditiveCurrents}
Let $\phi$ be an additive functorial interpretation of a polygraphic program~$\Pr$ and let~$t$ be a value with coarity $1$. Then the following equality holds in $\Nb$:
$$
\phi(t) \:=\: \sum_{\raisebox{-.75mm}{\smallfigeps{phi}}\in\Pr_2^C} \norm{t}_{\smallfigeps{phi}} \cdot c_{\smallfigeps{phi}}.
$$
\end{lem}

\proof
Let us prove this result by induction on the size of the $2$-path $t$. There is no degenerate value with coarity~$1$. If~$t$ is an elementary value with coarity~$1$, then~$t$ is a constructor $2$-cell~$\figeps{nil}$ with arity~$0$. Since~$\phi$ is additive, one has $\phi(\figeps{nil})=c_{\smallfigeps{nil}}$. Moreover,~$\norm{t}_{\smallfigeps{phi}}$ is~$1$ when $\figeps{phi}=\figeps{nil}$ holds and~$0$ otherwise, yielding the equality one seeks.

Now, let us fix a non-zero natural number~$N$ and assume that the result holds for every value with coarity~$1$ and size at most~$N$. Let us fix a value~$t$ with coarity~$1$ and size $N+1$. Then~$t$ admits a decomposition $t \:=\: \big( t_1\star_0\dots\star_0 t_n \big) \star_1 \figeps{phi}$, where~$\figeps{phi}$ is a constructor $2$-cell with arity~$n$ and each~$t_i$, $i\in\ens{1,\dots,n}$, is a value with coarity~$1$ and size at most~$N$. As a consequence, for every constructor $2$-cell~$\figeps{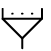}$, one has:
$$
\norm{t}_{\smallfigeps{phi-b}} 
\: = \:
\begin{cases}
\norm{t_1}_{\smallfigeps{phi-b}} + \cdots + \norm{t_n}_{\smallfigeps{phi-b}} + 1
	&\text{if $\figeps{phi-b}=\figeps{phi}$,} \\
\norm{t_1}_{\smallfigeps{phi-b}} + \cdots + \norm{t_n}_{\smallfigeps{phi-b}} 
	&\text{otherwise.}
\end{cases}
$$

\noindent Finally, one computes: 
\begin{align*}
\phi(t) \: 
&= \: \phi(\figeps{phi}) \circ \big( \phi(t_1)\times\dots\times\phi(t_n) \big) 
	&& \text{from the functorial relations of $\phi$,} \\
&= \: \phi(t_1) + \cdots + \phi(t_n) + c_{\smallfigeps{phi}} 
	&& \text{since $\phi$ is additive,} \\
&= \: \sum_{\raisebox{-.75mm}{\smallfigeps{phi-b}}\in\Pr_2^C} \left( \norm{t_1}_{\smallfigeps{phi-b}} + \cdots + \norm{t_n}_{\smallfigeps{phi-b}} \right) \cdot c_{\smallfigeps{phi-b}} + c_{\smallfigeps{phi}} 
	&& \text{by induction hypothesis} \\
&= \: \sum_{\raisebox{-.75mm}{\smallfigeps{phi-b}}\in\Pr_2^C} \norm{t}_{\smallfigeps{phi-b}} \cdot c_{\smallfigeps{phi-b}} 
	&&\text{from previous remark}.\rlap{\hbox to70 pt{\hfill\qEd}}
\end{align*}

\begin{prop}\label{Proposition:SizeValues}
Let $\phi$ be an additive functorial interpretation of a polygraphic program~$\Pr$. Then, for every value $t$ with coarity $1$, the inequalities $\norm{t}\leq \phi(t)\leq \gamma\norm{t}$ hold in $\Nb$. As a consequence, for every value $t$, one has $\nu(t) \leq \phi(t) \leq \gamma\nu(t)$, where~$\nu$ is the functorial interpretation introduced in Example~\ref{Example:FunctorialNorm}.
\end{prop}

\proof
Let us assume that $t$ is a value with coarity $1$. From Lemma~\ref{Lemma:AdditiveCurrents}, one has:
$$
\phi(t)
\:=\:
\sum_{\raisebox{-.75mm}{\smallfigeps{phi}}\in\Pr_2^C} \norm{t}_{\smallfigeps{phi}} \cdot c_{\smallfigeps{phi}}.
$$

\noindent By additivity of $\phi$ and by definition of $\gamma$, one has $1\leq c_{\smallfigeps{phi}}\leq\gamma$ for every constructor $2$-cell~$\figeps{phi}$. One concludes by using the following equality, that holds since $t$ is in $\mon{\Pr_2^C}$: 
$$
\norm{t}=\sum_{\raisebox{-.75mm}{\smallfigeps{phi}}\in\Pr_2^C} \norm{t}_{\smallfigeps{phi}}.
$$

\noindent When $t_1$, $\dots$, $t_n$ are values with coarity $1$ and when $t=t_1\star_0\dots\star_0 t_n$, one concludes thanks to the equalities $\phi(t) = \big( \phi(t_1), \dots,\phi(t_n) \big)$ and $\nu(t)= \big(\norm{t_1},\dots,\norm{t_n}\big)$.
\qed

\begin{lem}\label{Lemma:AdditiveCurrentsMax}
Let $\phi$ be an additive functorial interpretation of a polygraphic program $\Pr$. For every value $t$ with coarity $1$, the equality $\dr_{\phi} t=\phi(t)$ holds. As a consequence, for every value~$t$ with coarity~$n$, one has $\dr_{\phi} t=\mu_n\circ\phi(t)$. 
\end{lem}

\proof
Let us proceed by induction on the size of $t$. If $\figeps{nil}$ is a constructor $2$-cell with arity $0$, then the equality holds by definition of $\dr_{\phi}\figeps{nil}$. 

Now, let us fix a non-zero natural number $N$ and assume that the result holds for every value with coarity $1$ and size at most $N$. Let us consider a value $t$ with coarity $1$ and size $N+1$. One decomposes $t$ into $t=(t_1\star_0\dots\star_0 t_n)\star_1\figeps{phi}$, with $\figeps{phi}$ a constructor $2$-cell and where $t_i$ is a value with coarity $1$ and size at most $N$, for every $i\in\ens{1,\dots,n}$. Using the differential relations of $\dr_{\phi}$, one gets:
$$
\dr_{\phi} t \: = \: \max\ens{\: \dr_{\phi}(t_1), \:\dots, \: \dr_{\phi}(t_n), \: 
	\dr_{\phi} \figeps{phi} \: \big( \phi(t_1),\dots,\phi(t_n) \big) \: }.
$$

\noindent The definition of $\dr_{\phi}\figeps{phi}$ gives:
$$
\dr_{\phi} \figeps{phi} \: \big( \phi(t_1),\dots,\phi(t_n) \big) 
	\:=\: \max\ens{ \: \phi(t_1), \: \dots, \: \phi(t_n), \: \phi(\figeps{phi}) \big( \phi(t_1),\dots, \phi(t_n) \big) \: }.
$$

\noindent Since $\phi$ is additive, $\phi(\figeps{phi}) \big( \phi(t_1),\dots, \phi(t_n) \big)$ is greater than every $\phi(t_i)$, which is $\dr_{\phi}(t_i)$ by induction hypothesis applied to $t_i$. Thus one gets the following equality and uses the functorial relations of $\phi$ to conclude:
$$
\dr_{\phi} t \: = \: \phi(\figeps{phi}) \big( \phi(t_1),\dots, \phi(t_n) \big).
$$

\noindent Finally, let us consider a value $t$ with coarity $n$. One denotes by $(t_1,\dots,t_n)$ the family of values with coarity $1$ such that $t=t_1\star_0\dots\star_0 t_n$ holds. One invokes the differential relations of~$\dr_{\phi}$ to get the equality $\dr_{\phi}t = \max\big\{\:\dr_{\phi}(t_1),\:\dots,\:\dr_{\phi}(t_n)\:\big\}$. One uses the induction hypothesis on each $t_i$ and concludes, thanks to the functorial relations satisfied by $\phi$.
\qed

\begin{prop}\label{Proposition:AdditiveCurrentsMax}
Let $\phi$ be an additive functorial interpretation on a polygraphic program~$\Pr$. For every function $2$-cell $\figeps{fonction}$ and every value $t$ of type $s_1(\figeps{fonction})$, one has $\dr_{\phi}(t\star_1\figeps{fonction}) = \dr_{\phi}\figeps{fonction}\circ\phi(t)$.
\end{prop}

\proof
Let us assume that $\figeps{fonction}$ has valence $(m,n)$. One uses the differential relations of $\dr_{\phi}$ to produce:
$$
\dr_{\phi}(t\star_1\figeps{fonction}) \: = \: \max\ens{\dr_{\phi}t,\dr_{\phi}\figeps{fonction}\circ\phi(t)}.
$$

\noindent But, by definition of $\dr_{\phi}$, one has $\dr_{\phi}\figeps{fonction}\circ\phi(t) \geq \mu_m\circ\phi(t)$. There remains to use Lemma~\ref{Lemma:AdditiveCurrentsMax} on $t$ to get $\dr_{\phi}t=\mu_n\circ\phi(t)$. 
\qed

\begin{notation}
Let $\figeps{fonction}$ be a function $2$-cell with arity $m$ in a polygraphic program $\Pr$, equipped with an additive functorial interpretation $\phi$. Thereafter, we denote by $M_{\smallfigeps{fonction}}$ the map from~$\Nb^m$ to~$\Nb$ defined by:
$$
M_{\smallfigeps{fonction}}(x_1,\dots,x_m) \:=\: \dr_{\phi}\figeps{fonction} \: \big( \: \gamma x_1, \:\dots, \: \gamma x_m \: \big).
$$
\end{notation}

\noindent The next result uses the map $M_{\smallfigeps{fonction}}$ and the size of the initial arguments to bound the size of intermediate values produced during computations, hence of the arguments of potential recursive calls.

\begin{prop}\label{Proposition:SizeIntermediateValues}
Let $\Pr$ be a polygraphic program, equipped with an additive, compatible and conservative functorial interpretation $\phi$. Let $\figeps{fonction}$ be a function $2$-cell and let $t$ be a value of type $s_1\figeps{fonction}$. Then, for every $3$-path $F$ with source $t\star_1\figeps{fonction}$, the following inequality holds in~$\Nb$:
$$
\dr_{\phi}(t_2 F) \:\leq\: M_{\smallfigeps{fonction}} \circ \nu(t). 
$$
\end{prop}

\proof
The functorial interpretation $\phi$ is compatible and conservative: by Proposition~\ref{Proposition:ConservativeCurrents}, we know that $\dr_{\phi}(t_2 F)\leq \dr_{\phi}(t\star_1\figeps{fonction})$ holds. Since $\phi$ is additive, one may use Proposition~\ref{Proposition:AdditiveCurrentsMax} to produce the equality $\dr_{\phi}(t\star_1\figeps{fonction}) = \dr_{\phi}\figeps{fonction}\circ\phi(t)$. Furthermore, Proposition~\ref{Proposition:SizeValues} gives $\phi(t)\leq\gamma\nu(t)$: one argues that $\dr_{\phi}$ is monotone to conclude.
\qed

\begin{exa}
Applied to Example~\ref{Example:FusionSort}, Proposition~\ref{Proposition:SizeIntermediateValues} tells us that, given a list $t$, any intermediate value produced by the computation of the sorted list $\figeps{sort}(t)$ has its size bounded by $M_{\smallfigeps{sort}}(\norm{t})=\norm{t}$. This means that recursive calls made during this computation are applied to arguments of size at most $\norm{t}$.
\end{exa}

\subsection{Cartesian polygraphic interpretations and the size of structure computations}
\label{Subsection:StructureComplexity}

Here we bound the number of structure $3$-cells that can appear in a computation. For that, we consider polygraphic interpretations that take special values on structure $2$-cells.

\begin{defi}
Let $\Pr$ be a polygraphic program. A functorial interpretation $\phi$ of $\Pr$ is said to be \emph{cartesian} when the following conditions hold, for every $1$-cells $\xi$ and $\zeta$:
$$
\phi\left(\figeps{delta}_{\xi}\right)(x) \:=\: (x,x) 
\qquad\text{and}\qquad
\phi\left(\figeps{tau}_{\xi,\zeta}\right)(x,y) \:=\: (y,x).
$$

\noindent A polygraphic interpretation is \emph{cartesian} when its functorial part is cartesian and when its differential part sends every constructor and structure $2$-cell to zero.
\end{defi}

\begin{prop}\label{Proposition:CartesianCurrents}
If a functorial interpretation of a polygraphic program $\Pr$ is cartesian, then it is compatible with and conservative on all the structure $3$-cells. 
\end{prop}

\proof
Let $\phi$ be a cartesian functorial interpretation of a polygraphic program $\Pr$. We start by computing the values of $\phi$ and $\dr_{\phi}$ on the structure $2$-paths, by induction on their size. This way, one proves that the following equalities hold, for any $1$-path $u$ and $x\in\phi(u)$, any $1$-cell $\xi$ and $y\in\phi(\xi)$:
\begin{align*}
\phi\left(\figeps{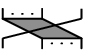}_{u,\xi}\right)(x,y) \:=\: (y,x) ,
&\qquad 
\phi\left(\reflectbox{\figeps{tau-n-1}}_{\xi,u}\right)(y,x) \:=\: (x,y), \\
\phi\left(\figeps{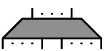}_u\right)(x) \:=\: (x,x),
&\qquad 
\phi\left(\figeps{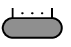}_u\right)(x) \:=\: \ast.
\end{align*}

\noindent Then, when $u=\ast$, all these $2$-paths are degenerate, so that they are sent on~$0$ by the differential interpretation~$\dr_{\phi}$. Now, when $u$ is non-degenerate, with $x=(x_1,\dots,x_n)$, one gets:
\begin{center}
$\dr_{\phi} \left(\figeps{tau-n-1}_{u,\xi} \right)(x,y)
\:=\: \max\ens{x_1,\dots,x_n,y}
\:=\: \dr_{\phi} \left(\reflectbox{\figeps{tau-n-1}}_{\xi,u}\right)(y,x)$,
\\
$\dr_{\phi} \left( \figeps{delta-n}_u \right) (x)
\:=\: \max\ens{x_1,\dots,x_n}
\:=\: \dr_{\phi} \left( \figeps{epsilon-n}_u \right) (x)$.
\end{center}

\noindent Now, we fix a $1$-path $u$, $1$-cells $\xi$, $\zeta$ and a constructor $2$-cell $\figeps{phi}:u\fl\xi$ in $\Pr$. Let us consider $x\in\phi(u)$ and $y\in\phi(\zeta)$ and check that the following equalities hold, yielding the compatibility of $\phi$ on structure $3$-cells:
$$
\phi\left( \raisebox{-1.25mm}{\figeps{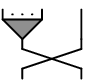}} \right) (x,y)
\:=\: (y,\phi(x))
\:=\: \phi\left( \raisebox{-1.25mm}{\figeps{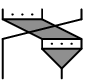}} \right) (x,y),
$$
$$
\phi\left( \raisebox{-1.25mm}{\reflectbox{\figeps{2-source-symetrie-gauche}}} \right) (y,x)
\:=\: (\phi(x),y)
\:=\: \phi\left( \raisebox{-1.25mm}{\reflectbox{\figeps{2-but-symetrie-gauche}}} \right) (y,x),
$$
$$
\phi\left( \raisebox{-1.25mm}{\figeps{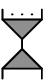}} \right) (x)
\:=\: (\phi(x),\phi(x))
\:=\: \phi\left( \raisebox{-1.25mm}{\figeps{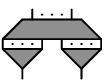}} \right) (x),
$$
$$
\phi\left( \raisebox{-1.25mm}{\figeps{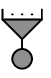}} \right) (x)
\:=\: \ast
\:=\: \phi\left(\figeps{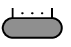}\right)(x).
$$
  With the same notations, we now check the conservativeness of $\phi$
  with the structure $3$-cells, \ie the compatibility of
  $\dr_{\phi}$ with them:
$$
\dr_{\phi}\left( \raisebox{-1.25mm}{\figeps{2-source-symetrie-gauche}} \right) (x,y)
\:=\: \max\ens{\dr_{\phi}\figeps{phi}(x),y}
\:\geq\: \dr_{\phi}\left( \raisebox{-1.25mm}{\figeps{2-but-symetrie-gauche}} \right) (x,y),
$$
$$
\dr_{\phi}\left( \raisebox{-1.25mm}{\reflectbox{\figeps{2-source-symetrie-gauche}}} \right) (y,x)
\:=\: \max\ens{\dr_{\phi}(\figeps{phi})(x),y}
\:\geq\: \dr_{\phi}\left( \raisebox{-1.25mm}{\reflectbox{\figeps{2-but-symetrie-gauche}}} \right) (y,x),
$$
$$
\dr_{\phi}\left( \raisebox{-1.25mm}{\figeps{2-source-duplication}} \right) 
\:=\: \dr_{\phi}(\figeps{phi})
\:=\: \dr_{\phi}\left( \raisebox{-1.25mm}{\figeps{2-but-duplication}} \right),
$$
$$
\dr_{\phi}\left( \raisebox{-1.25mm}{\figeps{2-source-effacement}} \right)
\:=\: \dr_{\phi}(\figeps{phi})
\:\geq\: \dr_{\phi}\left( \figeps{2-but-effacement} \right).\eqno{\qEd}
$$

\begin{defi}
Let $\phi$ be a functorial interpretation of a polygraphic program $\Pr$. We denote by $\dr_{\phi}^S$ and call \emph{structure differential interpretation generated by $\phi$} the differential interpretation of $\Pr$, over $\phi$ and into $(\Nb,+,0)$, that sends every constructor and function $2$-cell to zero and such that the following hold:
$$
\dr_{\phi}^S \figeps{tau} \:(x,y) \:=\: xy, \qquad
\dr_{\phi}^S \figeps{delta}\:(x) \:=\: x^2, \qquad
\dr_{\phi}^S \figeps{epsilon}\:(x) \:=\: x.
$$
\end{defi}

\begin{lem}\label{Lemma:StructureHeatCompatible}
Let $\phi$ be a functorial interpretation of a polygraphic program $\Pr$. If $\phi$ is both additive and cartesian, then $\dr_{\phi}^S$ is strictly compatible with all the structure $3$-cells of $\Pr$.
\end{lem}

\proof
We start by computing $\dr_{\phi}^S$ on the structure $2$-paths, by induction on their size:
\begin{center}
$
\dr_{\phi}^S\left( \figeps{tau-n-1} \right) (x_1,\dots,x_n,y) \: = \:
\dr_{\phi}^S\left( \reflectbox{\figeps{tau-n-1}} \right) (y,x_1,\dots,x_n) \: = \:
y \cdot \sum_{1\leq i\leq n} x_i,
$ \medskip \\
$
\dr_{\phi}^S\left( \figeps{delta-n} \right) (x_1,\dots,x_n) \: = \: \sum_{1\leq i\leq j\leq n} x_i\cdot x_j,
\qquad\qquad
\dr_{\phi}^S\left( \figeps{epsilon-n} \right) (x_1,\dots,x_n) \: = \: \sum_{1\leq i\leq n} x_i.
$
\end{center}

\noindent Now, let us fix a constructor $2$-cell $\figeps{phi}$ with arity $n$. Let us consider $x=(x_1,\dots,x_n)$ in $\phi(s_1\figeps{phi})$. Since~$\phi$ is additive, one notes that $\phi(\figeps{phi}) (x) \: > \: x_1 + \dots +x_n$ holds. Then, given a $y\in\Nb-\ens{0}$, one checks that the following strict inequalities hold in $\Nb-\ens{0}$:
\begin{align*}
\dr_{\phi}^S\left( \raisebox{-1.25mm}{\figeps{2-source-symetrie-gauche}} \right) (x,y)
\:=\: y \cdot \phi(\figeps{phi}) (x)
\:&>\: y \cdot \sum_{1\leq i\leq n} x_i
\:=\: \dr_{\phi}^S\left( \raisebox{-1.25mm}{\figeps{2-but-symetrie-gauche}} \right) (x,y), \\
\dr_{\phi}^S\left( \raisebox{-1.25mm}{\reflectbox{\figeps{2-source-symetrie-gauche}}} \right) (x,y)
\:=\: y \cdot \phi(\figeps{phi}) (x)
\:&>\: y \cdot \sum_{1\leq i\leq n} x_i
\:=\: \dr_{\phi}^S\left( \raisebox{-1.25mm}{\reflectbox{\figeps{2-but-symetrie-gauche}}} \right) (x,y), \\
\dr_{\phi}^S\left( \raisebox{-1.25mm}{\figeps{2-source-duplication}} \right)  (x)
\:=\: \left( \phi(\figeps{phi}) (x)\right)^2
\:&>\: \sum_{1\leq i\leq j\leq n} x_i\cdot x_j
\:=\: \dr_{\phi}^S\left( \raisebox{-1.25mm}{\figeps{2-but-duplication}} \right)  (x), \\
\dr_{\phi}^S\left( \raisebox{-1.25mm}{\figeps{2-source-effacement}} \right)  (x)
\:=\: \phi(\figeps{phi}) (x)
\:&>\: \sum_{1\leq i\leq n} x_i
\:=\: \dr_{\phi}^S\left( \figeps{2-but-effacement} \right)
(x).\rlap{\hbox to 91 pt{\hfil\qEd}}
\end{align*}

\noindent The following result gives sufficient conditions on a polygraphic interpretation such that one does not have to bother with the structure $3$-cells to prove termination.

\begin{prop}\label{Proposition:RelativeTerminationPrograms}
If a polygraphic program admits an additive and cartesian polygraphic interpretation that is strictly compatible with every computation $3$-cell, then it terminates.
\end{prop}

\proof
Let $(\phi,\dr)$ be a polygraphic interpretation with the required properties. One applies Proposition~\ref{Proposition:CartesianCurrents} to get the compatibility of $\phi$ with structure $3$-cells. Then Lemma~\ref{Lemma:StructureHeatCompatible} tells us that $(\phi,\dr_{\phi}^S)$ is strictly compatible with structure $3$-cells: hence Theorem~\ref{Theorem:Termination} yields termination of $\Pr_3^S$. 

Since $\dr$ sends every constructor and structure $2$-cell to zero, one has $\dr(s_2\alpha)=\dr(t_2\alpha)=0$ for every structure $3$-cell $\alpha$: thus $(\phi,\dr)$ is compatible with every structure $3$-cell and, by hypothesis, strictly compatible with every other $3$-cell. One applies Proposition~\ref{Proposition:RelativeTermination} to conclude.
\qed

\begin{defi}
Let $\Pr$ be a polygraphic program. One denotes by $K$ the maximum number of structure $2$-cells one finds in the targets of computation $3$-cells:
$$
K=\max\ens{\norm{t_2(\alpha)}_{\Pr_2^S},\alpha\in\Pr_3^R}.
$$

\noindent Let $\phi$ be an additive functorial interpretation of $\Pr$. For every function $2$-cell $\figeps{fonction}$ with arity~$m$, one defines~$S_{\smallfigeps{fonction}}$ as the map from~$\Nb^m$ to $\Nb$ given by:
$$
S_{\smallfigeps{fonction}} (x_1,\dots,x_m) \: = \: K\cdot M_{\smallfigeps{fonction}}^2(x_1,\dots,x_m).
$$
\end{defi}

\noindent The following lemma proves that, during a computation, if one applies a computation $3$-cell, then the structure heat increase is bounded by a polynomial in the size of the arguments.

\begin{lem}\label{Lemma:SizeManagement}
Let $\Pr$ be a polygraphic program, equipped with an additive, cartesian, compatible and conservative functorial interpretation $\phi$. Let $\figeps{fonction}$ be a function $2$-cell and $t$ be a value of type $s_1(\figeps{fonction})$. Let $f$ and $g$ be $2$-paths such that $t\star_1\figeps{fonction}$ reduces into $f$ which, in turn, reduces into $g$ by application of a computation $3$-cell~$\alpha$. Then, the following inequality holds in $\Zb$:
$$
\dr_{\phi}^S g - \dr_{\phi}^S f \:\leq\: S_{\smallfigeps{fonction}} \circ \nu(t).
$$
\end{lem}

\proof
Since $\phi$ is compatible and conservative, one can apply Proposition~\ref{Proposition:HeatIncrease} on the $3$-path from $f$ to~$g$, to get the following inequality:
$$
\dr_{\phi}^S g - \dr_{\phi}^S f \:\leq\:
\sum_{\raisebox{-.75mm}{\smallfigeps{fonction}}\in\Pr_2} \norm{t_2(\alpha)}_{\smallfigeps{fonction}} \cdot \dr_{\phi}^S \figeps{fonction} \: \big( \: \dr_{\phi}(f), \: \dots, \: \dr_{\phi}(f) \: \big).
$$

\noindent By definition of $\dr_{\phi}^S$, one has $\dr_{\phi}^S \figeps{fonction}=0$ except when $\figeps{fonction}$ is a structure $2$-cell. Thus one gets:
\begin{align*}
& \: \dr_{\phi}^S g - \dr_{\phi}^S f \\
\:\leq&\: 
\norm{t_2(\alpha)}_{\smallfigeps{tau}} \cdot \dr_{\phi}^S \figeps{tau} \: \big( \: \dr_{\phi}(f), \: \dr_{\phi}(f) \: \big) 
\:+\: \norm{t_2(\alpha)}_{\smallfigeps{delta}} \cdot \dr_{\phi}^S \figeps{delta} \: \big(  \dr_{\phi}(f)  \big) 
\:+\: \norm{t_2(\alpha)}_{\smallfigeps{epsilon}} \cdot \dr_{\phi}^S \figeps{epsilon} \: \big( \dr_{\phi}(f) \big) \\
=&\:
\norm{t_2(\alpha)}_{\smallfigeps{tau}} \cdot \big(\dr_{\phi}(f) \big)^2 
\:+\:  \norm{t_2(\alpha)}_{\smallfigeps{delta}} \cdot \big(\dr_{\phi}(f) \big)^2 
\:+\: \norm{t_2(\alpha)}_{\smallfigeps{epsilon}} \cdot \dr_{\phi}(f) \\
\leq&\:
\norm{t_2(\alpha)}_{\Pr_2^S} \cdot \big(\dr_{\phi}(f) \big)^2 \\
\leq&\:
K\cdot \big(\dr_{\phi}(f) \big)^2.
\end{align*}

\noindent Finally, we recall that $\phi$ is additive, compatible and
conservative: an application of Proposi\-tion
\ref{Proposition:SizeIntermediateValues} to the $3$-path with source
$t\star_1\figeps{fonction}$ and target $f$ yields $\dr_{\phi}(f)\leq
M_{\smallfigeps{fonction}} \circ\nu(t)$ and concludes the proof.  \qed

\begin{exa}
For the polygraphic program of Example~\ref{Example:FusionSort}, we
have $K=1$. The polynomials bounding the structure interpretation
increase after application of one of the computation $3$-cells of this
polygraphic program are:
$$
S_{\smallfigeps{sort}} (x) = x^2, \qquad
S_{\smallfigeps{split}} (x) = x^2,\qquad
S_{\smallfigeps{merge}} (x,y) = (x+y-1)^2.
$$
\end{exa}

\subsection{The size of computations}
\label{Subsection:TimeComplexity}

\begin{defi}
Let $\Pr$ be a polygraphic program, with an additive polygraphic
interpreta\-tion $(\phi,\dr)$. For every function $2$-cell
$\figeps{fonction}$ with arity $m$, one denotes by
$P_{\smallfigeps{fonction}}$ and by $Q_{\smallfigeps{fonction}}$ the
maps from $\Nb^m$ to~$\Nb$ defined by:
\begin{align*}
& P_{\smallfigeps{fonction}}(x_1,\dots,x_m) 
	\:=\: \dr\figeps{fonction} \: \big( \gamma x_1,\dots,\gamma x_m \big), \\
& Q_{\smallfigeps{fonction}} (x_1,\dots,x_m) 
	\:=\: P_{\smallfigeps{fonction}} (x_1,\dots,x_m) \cdot \left( 1 + S_{\smallfigeps{fonction}} (x_1,\dots,x_m) \right).
\end{align*}
\end{defi}

\noindent The following result bounds the number of computation $3$-cells in a reduction $3$-path, with respect to the size of the arguments.

\begin{prop}\label{Proposition:SizeComputations}
Let $\Pr$ be a polygraphic program, equipped with an additive and cartesian polygraphic interpretation $(\phi,\dr)$ which is strictly compatible with every computation $3$-cell. Let $\figeps{fonction}$ be a function $2$-cell and $t$ be a value of type $s_1(\figeps{fonction})$. Then, for every $3$-path $F$ with source $t\star_1\figeps{fonction}$, the following inequality holds:
$$
\tnorm{F}_{\Pr_3^R} \:\leq\: P_{\smallfigeps{fonction}} \circ \nu(t).
$$
\end{prop}

\proof
If $F$ is degenerate, then $\tnorm{F}_{\Pr_3^R}=0$ holds. Otherwise, the $3$-path~$F$ decomposes this way:
$$
F \: = \: H_0 \star_2 G_1 \star_2 H_1 \star_2 G_2 \star_2 \cdots \star_2 G_k\star_2 H_k,
$$

\noindent where each $G_i$ is elementary in $\mon{\Pr_3^R}$ and each $H_j$ lives in $\mon{\Pr_3^S}$. Hence $\tnorm{F}_{\Pr_3^R}=k$. Since the polygraphic interpretation is cartesian, it is compatible with every structure $3$-cell, so that one has $\dr(s_2 H_j)\geq\dr(t_2 H_j)$, for every $j\in\ens{0,\dots,k}$. Since it is also strictly compatible with every computation $3$-cell, one applies Proposition~\ref{Proposition:CompatibleHeats} to get the following chain of (in)equalities, for every $i\in\ens{0,\dots,k-1}$:
$$
\dr(s_2 H_i) \geq \dr(t_2 H_i) = \dr (s_2 G_i) > \dr (t_2 G_i) = \dr(s_2 H_{i+1}).
$$

\noindent By induction on $i$, one proves the following chain of (in)equalities:
$$
\dr(t\star_1\figeps{fonction}) = \dr (s_2 G_1) > \dr (s_2 G_2) > \cdots > \dr (s_2 G_k) > \dr (t_2 G_k).
$$

\noindent Furthermore we have $\dr(t_2 G_k)\geq 0$ and, consequently:
$$
\tnorm{F}_{\Pr_3^R} \:\leq\: \dr(t\star_1\figeps{fonction}).
$$

\noindent Finally, let us bound $\dr(t\star_1\figeps{fonction})$, which is equal to $\dr\figeps{fonction} \circ \phi(t) + \dr t$, thanks to the differential relations of~$\dr$. But $(\phi,\dr)$ is cartesian, yielding $\dr t=0$, and Proposition~\ref{Proposition:SizeValues} tells us that $\phi(t)\leq \gamma\nu(t)$ holds.  One uses the definition of $P_{\smallfigeps{fonction}}$ to conclude.
\qed

\begin{prop}\label{Proposition:TotalSizeComputations}
Let $\Pr$ be a polygraphic program, equipped with an additive and cartesian polygraphic interpretation $(\phi,\dr)$ which is strictly compatible with and conservative on every computation $3$-cells. Let~$\figeps{fonction}$ be a function $2$-cell and let $t$ be a value of type $s_1\figeps{fonction}$. Then, for every $3$-path $F$ with source $t\star_1\figeps{fonction}$, the following inequality holds:
$$
\tnorm{F} \:\leq\: Q_{\smallfigeps{fonction}} \circ \nu(t).
$$
\end{prop}

\proof
If $\tnorm{F}=0$, then the inequality does hold. Otherwise, there exists a $3$-cell that we can apply to the starting $2$-path $t\star_1\figeps{fonction}$; moreover, this is a computation $3$-cell since no structure $3$-cell can be applied to such a $2$-path. Hence the $3$-path~$F$ decomposes this way:
$$
F \: = \: G_1 \star_2 H_1 \star_2 G_2 \star_2 \cdots \star_2 G_k\star_2 H_k,
$$

\noindent where each $G_i$ is elementary in $\mon{\Pr_3^R}$ and each $H_j$ is in $\mon{\Pr_3^S}$. As a consequence, we have: 
$$
\tnorm{F} \:=\: k+\tnorm{H_1}+\dots+\tnorm{H_k}.
$$

\noindent Furthermore $k=\tnorm{F}_{\Pr_3^R}$ holds and, thus, so does $k\leq P_{\smallfigeps{fonction}}\circ\nu(t)$ thanks to Proposition~\ref{Proposition:SizeComputations}. We prove that the following inequality holds to conclude:
$$
\tnorm{H_1}+\dots+\tnorm{H_k} \:\leq\: k\cdot \big( S_{\smallfigeps{fonction}} \circ\nu(t) \big).
$$

\noindent Towards this goal, let us fix an $i\in\ens{1,\dots,k}$. Since $\dr_{\phi}^S$ is strictly compatible with every structure $3$-cell, one gets from Proposition~\ref{Proposition:CompatibleHeats}:
$$
\tnorm{H_i} + \dr_{\phi}^S(t_2 H_i) \:\leq\: \dr_{\phi}^S(s_2 H_i).
$$

\noindent Furthermore, from Lemma~\ref{Lemma:SizeManagement}, one knows that the following inequality holds:
$$
\dr_{\phi}^S(t_2 G_i) \: \leq \: \dr_{\phi}^S(s_2 G_i) + S_{\smallfigeps{fonction}}\circ\nu(t).
$$

\noindent Since $t_2 G_i=s_2 H_i$ holds, one has:
$$
\tnorm{H_i} + \dr_{\phi}^S(t_2 H_i) \: \leq\:  \dr_{\phi}^S(s_2 G_i) + S_{\smallfigeps{fonction}}\circ\nu(t).
$$

\noindent Or, written differently:
$$
\tnorm{H_i} \: \leq \: \dr_{\phi}^S(s_2 G_i) - \dr_{\phi}^S(t_2 H_i) + S_{\smallfigeps{fonction}}\circ\nu(t).
$$

\noindent One sums this family of $k$ inequalities, one for every $i$ in $\ens{1,\dots,k}$, to produce:
$$
\tnorm{H_1}+\dots+\tnorm{H_k} \: \leq \: 
	\sum_{i=1}^k \dr_{\phi}^S(s_2 G_i) 
	- \sum_{i=1}^k \dr_{\phi}^S(t_2 H_i)
	+ k \cdot S_{\smallfigeps{fonction}}\circ\nu(t).
$$

\noindent By hypothesis, one has $s_2 G_1=t\star_1\figeps{phi}$, $t_2 H_k=t_2 F$ and, for every $i\in\ens{1,\dots,k}$, $t_2 H_i=s_2 G_{i+1}$, so that the following inequality holds:
$$
\tnorm{H_1}+\dots+\tnorm{H_k} \:\leq\: \dr_{\phi}^S(s_2 F) - \dr_{\phi}^S(t_2 F) + k\cdot S_{\smallfigeps{fonction}}\circ\nu(t).
$$

\noindent Finally, one argues that both $\dr_{\phi}^S(t\star_1\figeps{fonction})=0$ and $\dr_{\phi}^S(t_2 F)\geq 0$ hold by definition of $\dr_{\phi}^S$.
\qed

\begin{exa}
Let us compute these bounding maps for the fusion sort function $2$-cell $\figeps{sort}$ of the polygraphic program~$\Fr$:
$$
P_{\smallfigeps{sort}} (2x+1) \:=\: 2x^2+1
\qquad\text{and}\qquad
Q_{\smallfigeps{sort}} (2x+1) \:=\: (2x^2+1) \cdot \big( 1+(2x+1)^2 \big).
$$

\noindent Let us fix a list $[i_1;\dots;i_n]$ of natural numbers. One can check that, in $\Fr$, this list is represented by a $2$-path $t$ such that $\phi(t)=\norm{t}=2n+1$. The polynomial $P_{\smallfigeps{sort}}$ tells us that, during the computation of the sorted list~$\sem{\figeps{sort}}(t)$, there will be at most $2n^2+1$ applications of computation $3$-cells. The polynomial~$Q_{\smallfigeps{sort}}$ bounds the total number of $3$-cells of any type. 

For example, when $n$ is $2$, one computes $\sem{\figeps{sort}}(t)$ by building a $3$-path of size at most $Q_{\smallfigeps{sort}}(5)=234$, containing no more than $P_{\smallfigeps{sort}}(5)=9$ computation $3$-cells. One can check that the $3$-path presented in Example~\ref{Example:Reduction} is (way) below these bounds: it is made of seven $3$-cells, six of which are of the computation kind.
\end{exa}

\subsection{Polygraphic programs and polynomial-time functions}
\label{Subsection:PTIME}

\begin{defi}
Let $\Pr$ be a polygraphic program. A differential interpretation~$\dr$ of $\Pr$ is \emph{polynomial} when, for every function $2$-cell $\figeps{fonction}$, the map $\dr\figeps{fonction}$ is bounded by a polynomial. A functorial interpretation~$\phi$ of $\Pr$ is \emph{polynomial} when $\dr_{\phi}$ is. A polygraphic interpretation is \emph{polynomial} when both its functorial part and differential part are.

We denote by $\Pbf$ the set of polygraphic programs which are confluent and complete and which admit an additive, cartesian and polynomial polygraphic interpretation that is conservative on and strictly compatible with their computation $3$-cells.
\end{defi}

\begin{exa} 
As a consequence of previous results, the two polygraphic programs $\Dr$, computing euclidean division, and $\Fr$, computing the fusion sort of lists, are in $\Pbf$. 
\end{exa}

\begin{defi}
Let us denote by $\Nr$ the polygraphic program with the following cells:
\begin{enumerate}[(1)]

\item It has one $1$-cell $\mathtt{n}$.

\item Its $2$-cells are the three possible structure $2$-cells plus: 
\begin{enumerate}
\item Constructor $2$-cells: $\figeps{nil}$ for zero and $\figeps{succ}$ for the successor.
\item Function $2$-cells: $\figeps{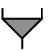}$ for addition and $\figeps{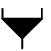}$ for multiplication.
\end{enumerate}

\item Its $3$-cells are the eight structure $3$-cells plus the following computation $3$-cells:
\end{enumerate}
\begin{center}\input{3-cellules-polynomes.pstex_t}\end{center}
\end{defi}

\begin{prop}\label{Proposition:Polynomials}
The polygraphic program $\Nr$ is in $\Pbf$ and it computes the addition and multiplication of natural numbers.
\end{prop}

\proof
The polygraphic program $\Nr$ is orthogonal, hence locally confluent, and complete. Furthermore, the following hold:
$$
\sem{\mathtt{n}}\simeq\Nb, \qquad 
\sem{\figeps{add}} \left(\ul{m},\ul{n}\right) = \ul{m+n}, \qquad 
\sem{\figeps{mult}} \left(\ul{m},\ul{n}\right)=\ul{mn}.
$$

\noindent Then, one checks that the following polygraphic interpretation has all the required properties:
$$
\phi(\mathtt{n})=\Nb-\ens{0}, \qquad c_{\smallfigeps{nil}}=c_{\smallfigeps{succ}}=1, \qquad \phi(\figeps{add})(x,y)=x+y, \qquad \phi(\figeps{mult})(x,y)=xy,
$$
$$
\dr\figeps{add}(x,y)=x \qquad\text{and}\qquad \dr\figeps{mult}(x,y)=(x+1)y.\eqno{\qEd}
$$

\begin{rem}
  So $\Nr$ computes addition and multiplication of natural numbers. As
  we have seen, it also computes duplication and permutation on
  them. As a consequence, for every polynomial $P$ in $\Nb[x]$, one
  can choose a $2$-path $\raisebox{-1.25mm}{\begin{picture}(0,0)%
\includegraphics{poly.pstex}%
\end{picture}%
\setlength{\unitlength}{4144sp}%
\begingroup\makeatletter\ifx\SetFigFont\undefined%
\gdef\SetFigFont#1#2#3#4#5{%
  \reset@font\fontsize{#1}{#2pt}%
  \fontfamily{#3}\fontseries{#4}\fontshape{#5}%
  \selectfont}%
\fi\endgroup%
\begin{picture}(153,204)(-435,287)
\put(-359,355){\makebox(0,0)[b]{\smash{{\SetFigFont{6}{7.2}{\rmdefault}{\mddefault}{\updefault}{\color[rgb]{0,0,0}$P$}%
}}}}
\end{picture}%
}$ in $\Nr$ such that
  $\sem{\raisebox{-1.25mm}{\begin{picture}(0,0)%
\includegraphics{poly.pstex}%
\end{picture}%
\setlength{\unitlength}{4144sp}%
\begingroup\makeatletter\ifx\SetFigFont\undefined%
\gdef\SetFigFont#1#2#3#4#5{%
  \reset@font\fontsize{#1}{#2pt}%
  \fontfamily{#3}\fontseries{#4}\fontshape{#5}%
  \selectfont}%
\fi\endgroup%
\begin{picture}(153,204)(-435,287)
\put(-359,355){\makebox(0,0)[b]{\smash{{\SetFigFont{6}{7.2}{\rmdefault}{\mddefault}{\updefault}{\color[rgb]{0,0,0}$P$}%
}}}}
\end{picture}%
}}$ is $P$. Moreover, by induction, one proves
  that $\phi(\raisebox{-1.25mm}{\begin{picture}(0,0)%
\includegraphics{poly.pstex}%
\end{picture}%
\setlength{\unitlength}{4144sp}%
\begingroup\makeatletter\ifx\SetFigFont\undefined%
\gdef\SetFigFont#1#2#3#4#5{%
  \reset@font\fontsize{#1}{#2pt}%
  \fontfamily{#3}\fontseries{#4}\fontshape{#5}%
  \selectfont}%
\fi\endgroup%
\begin{picture}(153,204)(-435,287)
\put(-359,355){\makebox(0,0)[b]{\smash{{\SetFigFont{6}{7.2}{\rmdefault}{\mddefault}{\updefault}{\color[rgb]{0,0,0}$P$}%
}}}}
\end{picture}%
})=P$ and that $\dr\raisebox{-1.25mm}{\begin{picture}(0,0)%
\includegraphics{poly.pstex}%
\end{picture}%
\setlength{\unitlength}{4144sp}%
\begingroup\makeatletter\ifx\SetFigFont\undefined%
\gdef\SetFigFont#1#2#3#4#5{%
  \reset@font\fontsize{#1}{#2pt}%
  \fontfamily{#3}\fontseries{#4}\fontshape{#5}%
  \selectfont}%
\fi\endgroup%
\begin{picture}(153,204)(-435,287)
\put(-359,355){\makebox(0,0)[b]{\smash{{\SetFigFont{6}{7.2}{\rmdefault}{\mddefault}{\updefault}{\color[rgb]{0,0,0}$P$}%
}}}}
\end{picture}%
}$ is bounded
  by a polynomial in $\Nb[x]$.
\end{rem}

\begin{thm}\label{Theorem:PTIME}
The polygraphic programs of $\Pbf$ compute exactly the \Ptime functions.
\end{thm}

\proof
  The fact that a function computed by a polygraphic program in $\Pbf$
  is in \Ptime is a consequence of the results of
  Proposition~\ref{Proposition:TotalSizeComputations}. Indeed, it
  proves that the size of any computation of~$\sem{\figeps{fonction}}$
  is bounded by~$Q_{\smallfigeps{fonction}}$ applied to the size of
  the arguments: from the polynomial assumption and the definition
  of~$Q_{\smallfigeps{fonction}}$, this map is itself bounded by a
  polynomial. Moreover each $3$-cell application modifies only
  finitely many $2$-cells: hence the sizes of the $2$-paths remain
  polynomial all along the computation. Furthermore, any step of
  computation can be done in polynomial time with respect to the size
  of the current $2$-path. Indeed, it corresponds to finding a pattern
  and, then, replace it by another one: it is just a reordering of
  some pointers with a finite number of memory allocations. So, the
  computation involves a polynomial number of steps, each of which can
  be performed in polynomial time. Thus, the normalization process can
  be done in polynomial time.

  Conversely, let $f:\mon{\Sigma}\fl\mon{\Sigma}$ be a function of
  class \Ptime. This means that there exists a Turing machine
  $\Mr=(\Sigma,Q,q_0,q_f,\delta)$ and a polynomial~$P$ in $\Nb[x]$
  such that the machine $\Mr$ computes $f$ and, for any word~$w$ of
  length $n$ in $\mon{\Sigma}$, the number of transition steps
  required by $\Mr$ to compute $f(w)$ is bounded by~$P(n)$. We extend
  the polygraphic program~$\Nr$ into~$\Pr(\Mr,P)$, by adding the
  following extra cells, adapted from the ones of the polygraphic
  Turing machine $\Pr(\Mr)$ used in the proof of
  Theorem~\ref{Theorem:TuringComplete}, in order to use~$P$ as a
  clock:

\begin{enumerate}[(1)]

\item An extra $1$-cell $\mathtt{w}$.

\item Extra $2$-cells include the five new structure $2$-cells plus:

\begin{enumerate}[(a)]

\item Constructor $2$-cells: the empty word
  $\figeps{nil}:\ast\dfl\mathtt{w}$ and each letter
  $\raisebox{-1.25mm}{\begin{picture}(0,0)%
\includegraphics{a.pstex}%
\end{picture}%
\setlength{\unitlength}{4144sp}%
\begingroup\makeatletter\ifx\SetFigFont\undefined%
\gdef\SetFigFont#1#2#3#4#5{%
  \reset@font\fontsize{#1}{#2pt}%
  \fontfamily{#3}\fontseries{#4}\fontshape{#5}%
  \selectfont}%
\fi\endgroup%
\begin{picture}(123,204)(344,557)
\put(406,637){\makebox(0,0)[b]{\smash{{\SetFigFont{6}{7.2}{\rmdefault}{\mddefault}{\updefault}{\color[rgb]{0,0,0}$a$}%
}}}}
\end{picture}%
}:\mathtt{w}\dfl\mathtt{w}$ of $\Sigma$.

\item Function $2$-cells: the main
  $\figeps{sort}:\mathtt{w}\dfl\mathtt{w}$ for $f$, plus the modified
  $\raisebox{-0.5mm}{\raisebox{-1.25mm}{\input{clock-step.pstex_t}}}$, $q\in Q$ and
  $a\in\overline{\Sigma}$, now from
  $\mathtt{n}\star_0\mathtt{w}\star_0\mathtt{w}$ to $\mathtt{w}$, plus
  an extra size function $\figeps{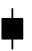}:\mathtt{w}\dfl\mathtt{n}$.

\end{enumerate}

\item Extra $3$-cells include the new structure ones plus:

\begin{enumerate}[(a)]
\item The computation $3$-cells for the auxiliary function $\figeps{size}$:
\begin{center}\input{3-cellules-size.pstex_t}\end{center}

\item Timed versions of the computation $3$-cells for the Turing machine:

\smallskip
\raisebox{-1.25mm}{\input{3-cellules-turing-clock.pstex_t}}
\end{enumerate}
\end{enumerate}

\noindent One checks that $\Pr(\Mr,P)$ is orthogonal and complete. We equip it with the polygraphic interpretation based on the one defined on~$\Nr$ in the proof of Proposition~\ref{Proposition:Polynomials}, extended with the following values:
$$
c_{\smallfigeps{nil}} = c_{\scalebox{0.66}{\begin{picture}(0,0)%
\includegraphics{a.pstex}%
\end{picture}%
\setlength{\unitlength}{4144sp}%
\begingroup\makeatletter\ifx\SetFigFont\undefined%
\gdef\SetFigFont#1#2#3#4#5{%
  \reset@font\fontsize{#1}{#2pt}%
  \fontfamily{#3}\fontseries{#4}\fontshape{#5}%
  \selectfont}%
\fi\endgroup%
\begin{picture}(123,204)(344,557)
\put(406,637){\makebox(0,0)[b]{\smash{{\SetFigFont{6}{7.2}{\rmdefault}{\mddefault}{\updefault}{\color[rgb]{0,0,0}$a$}%
}}}}
\end{picture}%
}}=1,
$$
$$
\qquad \phi(\figeps{size})(x)=x, \qquad \phi(\raisebox{-0.5mm}{\raisebox{-1.25mm}{\input{clock-step.pstex_t}}})(x,y,z)=x+y+z, \qquad \phi(\figeps{sort})(x) = P(x) + x + 1,
$$
$$
\dr\figeps{size}(x) = \dr\raisebox{-0.5mm}{\raisebox{-1.25mm}{\input{clock-step.pstex_t}}}(x,y,z) = x, \qquad \dr\figeps{sort}(x) = \dr\raisebox{-1.25mm}{\begin{picture}(0,0)%
\includegraphics{poly.pstex}%
\end{picture}%
\setlength{\unitlength}{4144sp}%
\begingroup\makeatletter\ifx\SetFigFont\undefined%
\gdef\SetFigFont#1#2#3#4#5{%
  \reset@font\fontsize{#1}{#2pt}%
  \fontfamily{#3}\fontseries{#4}\fontshape{#5}%
  \selectfont}%
\fi\endgroup%
\begin{picture}(153,204)(-435,287)
\put(-359,355){\makebox(0,0)[b]{\smash{{\SetFigFont{6}{7.2}{\rmdefault}{\mddefault}{\updefault}{\color[rgb]{0,0,0}$P$}%
}}}}
\end{picture}%
}(x)+P(x)+x+1.
$$

\noindent One checks that this polygraphic interpretation is additive,
cartesian, polynomial, compati\-ble with and conservative on all the
computation $3$-cells. Hence, $\Pr(\Mr,P)$ is a polygraphic program in
$\Pbf$. Furthermore, one has $\sem{\mathtt{n}}\simeq\Nb$ and
$\sem{\mathtt{w}}\simeq\mon{\Sigma}$. We also note that, among
functions computed by $\Pr(\Mr,P)$, one proves that
$\sem{\figeps{size}}:\sem{\mathtt{w}}\fl\sem{\mathtt{n}}$ is the
length function.

The four middle families of computation $3$-cells of $\Nr$ are once
again in bijection with the rules defining the transition relation of
the Turing machine $\Mr$. Hence, the configuration $(q,a,w_l,w_r)$
reduces into $(q',a',w'_l,w'_r)$ in $k\in\Nb$ steps if and only if,
for any $n\geq k$, one has:
$$
\left( \ul{n}\star_0 \ul{w_l} \star_0 \ul{w_r} \right) \star_1 \raisebox{-0.5mm}{\raisebox{-1.25mm}{\input{clock-step.pstex_t}}}
\quad \tfl \quad
\left( \ul{n-k} \star_0 \ul{w'_l} \star_0 \ul{w'_r} \right) \star_1 \raisebox{-0.5mm}{\raisebox{-1.25mm}{\input{clock-step-q-a-prime.pstex_t}}}.
$$

\noindent Finally, let us fix a word $w$ of length $n$ in
$\mon{\Sigma}$. The Turing machine computes~$f$, so that
$(q_0,\sharp,e,w)$ reduces into a unique configuration
$(q_f,a,v,f(w))$, after a finite number $k$ of transition steps. Then
we check the following chain of equalities, yielding
$\sem{\figeps{sort}}=f$:
$$
\sem{\figeps{sort}} \left( \ul{w} \right) 
	\:=\: \sem{ \raisebox{-0.5mm}{\raisebox{-1.25mm}{\input{clock-step-q0-a.pstex_t}}} } \left( \ul{P(n)} \star_0 \figeps{nil} \star_0 \ul{w} \right)
	\:=\: \sem{ \raisebox{-0.5mm}{\raisebox{-1.25mm}{\input{clock-step-qf-a.pstex_t}}} } \left( \ul{P(n)-k} \star_0 \ul{v} \star_0 \ul{f(w)} \right)
	\:=\: \ul{f(w)}.\eqno{\qEd}
$$

\section*{Future directions}
\label{Section:Conclusion}

\subsection*{Polygraphic programs} 
The definition we have chosen for this study stays close to the one of
first-order functional programs. We shall explore generalization along
different directions.

We think that an important research trail concerns the understanding
of the algebraic properties of the \emph{if-then-else} construction in
polygraphic terms. Towards this goal, we want to describe strategies
as sets of $4$-dimensional cells. The $3$-paths will contain all the
computational paths one can build when there is no fixed evaluation
strategy, while the strategies and conditions will be represented by
the $4$-paths, seen as normalization processes of $3$-paths. In
particular, this setting shall allow us to internalize the test used
to compute the merge function in the fusion sort algorithm, but also
to describe conditional or probabilistic rewriting systems.

On another point, in the polygraphs we consider here, we have fixed a
sublayer made of permutations, duplications and erasers, together with
natural polygraphic interpretations for them. However, one can see
them as a special kind of function $2$-cells. Thus, we shall define a
notion of hierarchical programs, where one builds functions level
after level, giving complexity bounds for them modulo the previously
defined functions. However, this does not prevent us to build modules
that a programmer can freely use as sublayers, without bothering with
the complexity of their functions: for example, a module that
describes the evaluation and coevaluation. We think of this module
system as a first possibility to integrate polymorphism into the
polygraphic setting.

Removing duplication and erasure from the standard definition means
that one moves from a cartesian setting to a monoidal one. According
to a variant of André Joyal's paradox~\cite{LambekScott86}, this is
necessary to describe functions such as linear maps on
finite-dimensio\-nal vector spaces. Thus, one should be able to compute,
for example, algebraic cooperations, such as the ones found in
Jean-Louis Loday's generalized bialgebras~\cite{Loday06}, or
automor\-phisms of $\Cb^n$, such as the universal Deutsch
gate~\cite{NielsenChuang00} of quantum circuits.

Going further, at this step, there will be no reason anymore to
consider constructor $2$-cells with one output only or values with no
output. This way, one could consider algorithms computing, for
example, on braids or knots. However, this also suggests to change our
notion of function $2$-cells to some kind of "polygraphic context", a
notion of $2$-path with holes whose algebraic structure has yet to be
understood. In particular, this is the second solution we think of to
describe polymorphic types and functions.

For all this research, we shall consider a more abstract definition of
polygraphs: they are special higher-dimensional categories, namely the
free ones. This formulation, though leading to a steeper learning
curve, shall provide enlightenments about the possibilities one has
when one wants to extend the setting. But, more importantly, this will
make easier the adaptation of tools from algebra for program analysis.

\subsection*{Analysis tools} 
In future work, we shall use other possibilities provided by
polygraphic interpretations, together with other algebraic tools, to
study the computational properties of polygraphs.

We restricted interpretations to be polynomials with integer
coefficients. This is close to the tools considered
in~\cite{BonfanteCichonMarionTouzet01}. Following this last paper, a
straightforward characteri\-za\-tion of exponential-time (resp. doubly
exponential-time) can be done by considering linear (resp. polynomial)
interpretations for constructors, instead of additive ones. However,
some studies are much more promising. First, to turn to polynomials
over reals give some procedures to build interpretations
(see~\cite{BonfanteMarionMoyenPechoux05}) via Alfred Tarski's
decidability~\cite{Tarski51}. Second, we plan to consider differential
interpretations with values in multisets (instead of natural numbers),
to characterize polynomial-space computations.

For each generalization of the notion of polygraphic program, such as
the ones mentioned earlier, we shall adapt polygraphic interpretations
in consequence. We think that, if these generalizations are done in an
elegant way, this task will be easier. For example, if one considers
"symmetric" values, \ie values with inputs, one can use a third part
of polygraphic interpretations we have not used here: ascending
currents, described by a contravariant functorial part, such as in the
original definition~\cite{Guiraud06jpaa}.

As pointed earlier, polygraphs are higher
dimensional-categories. Philippe Malbos and the second author are
currently adapting the finite derivation criterion of Craig
Squier \cite{Squier94} to them, as was done before for
$1$-categories~\cite{Malbos}. We think that this will lead us to a
computable necessary condition to ensure that a function admits a
finite, convergent polygraphic program that computes it.

The same collaboration has more long-term aims: using tools from
homological algebra for program analysis. For example, the functorial
and differential interpretations are special cases of, respectively,
left modules over the $2$-category of $2$-paths (or bimodules, when
there are ascending currents) and derivations of this same
$2$-category into the given module. Moreover, a well-chosen cohomology
theory yields, in particular, information on derivations: thus, one
can hope to get new tools such as negative results about the fact that
a given algorithm lives in a given complexity class.

\subsection*{Cat}
The main concrete objective of this project is to develop a new
programming language, codenamed Cat. In this setting, one will build a
program as a polygraph, while using the algebraic analysis tools we
provide to produce certificates that guarantee several properties of
the code, such as grammatical ones, computational ones or semantical
ones. As in Caml~\cite{Caml}, a Cat program will have two aspects: an
implementation and an interface.

In the implementation, one builds the code, describing the cells and
assembling them to build paths, \ie building the data types, the
functions, the computation rules and the evaluation strategies. Thanks
to the dual nature of polygraphs, one shall be able to perform this
using an environment that is either totally graphical, totally
syntactical or some hybrid possibility between those.

The interface part contains all the information the programmer can
prove on its code, in the form of certificates. These guaranteed
properties will range from type information, as in Caml, to
polygraphic interpretations proving termination or giving complexity
bounds, to proofs of semantical properties in the form of polygraphic
three-dimensional proofs~\cite{Guiraud06apal}. For all these
certificates, we shall propose assistants, with tactics that
automatize the simpler tasks and leave the programmer concentrate on
the harder parts.

Finally, given such a polygraphic program, the question of evaluation
arises. One can think of several solutions, whose respective
difficulty ranges from "feasible" to "science-fic\-tion": first, a
compiler or an interpreter into some existing language, such as
Tom~\cite{Tom}, a task that has already been started; then, a
distributed execution where each $2$-cell is translated into a
process, whose behaviour is described by the corresponding $3$-cells;
finally, concrete electronic chips dedicated to polygraphic
computation.

\bibliographystyle{alpha}
\bibliography{bibliographie}
\end{document}